\documentclass[fleqn,usenatbib]{mnras}

\usepackage{newtxtext}
\usepackage{newtxmath}

\usepackage[T1]{fontenc}
\usepackage{cancel}

\DeclareRobustCommand{\VAN}[3]{#2}
\let\VANthebibliography\thebibliography
\def\thebibliography{\DeclareRobustCommand{\VAN}[3]{##3}\VANthebibliography}

\usepackage{graphicx}	
\graphicspath{ {./figures/} }
\usepackage{amsmath}	
\usepackage{upgreek}    
\usepackage{soul}       

\def\gtorder{\mathrel{\raise.3ex\hbox{$>$}\mkern-14mu
    \lower0.6ex\hbox{$\sim$}}}
\def\ltorder{\mathrel{\raise.3ex\hbox{$<$}\mkern-14mu
    \lower0.6ex\hbox{$\sim$}}}

\DeclareSymbolFont{CMletters}{OML}{cmm}{m}{it}
\DeclareMathSymbol{v}{\mathord}{CMletters}{`v}
\DeclareMathOperator{\sech}{sech}

\title[Bars in Spinning Halos]{Stellar Bars in Spinning Halos:\\ Delayed Buckling and Absence of Slowdown} 

\author[X. Li et al.]{
Xingchen Li,$^{1}$\thanks{E-mail: xingchen.li@uky.edu}
Isaac Shlosman,$^{1,2}$\thanks{E-mail: isaac.shlosman@uky.edu}
Clayton Heller,$^{3}$
Daniel Pfenniger$^{4}$
\\
$^{1}$\rm Department of Physics \& Astronomy, University of Kentucky, Lexington, KY 40506, USA\\
$^{2}$\rm Theoretical Astrophysics, School of Sciences, Osaka University, Osaka 560-0043, Japan\\
$^{3}$\rm Department of Physics \& Astronomy, Georgia Southern University, Statesboro, GA 30460, USA\\
$^{4}$\rm  University of Geneva, Geneva Observatory, ch. Pegasi 51, 1290 Versoix, Switzerland
}

\date{Accepted XXX. Received YYY; in original form ZZZ}

\pubyear{2022}
\begin{document}
\label{firstpage}
\pagerange{\pageref{firstpage}--\pageref{lastpage}}
\maketitle

\begin{abstract}
We use numerical simulations to analyze the stellar bar evolution in spinning dark matter (DM) halos. Previous works have shown the halo spin has a substantial effect on bar evolution and can lead to bar dissolution following the vertical buckling instability. We invoke the DM spin sequence, $\uplambda = 0 - 0.09$, and study the effect of DM density along this $\uplambda$-sequence by varying compactness of DM halo. We find that (1) varying DM density has a profound effect on bar evolution along $\uplambda$-sequence; (2) For $\uplambda\gtorder 0.045$, the buckling has been delayed progressively; (3) Stellar bars remain near maximal strength, and their amplitude plateau stage extends over $0.7-5$\,Gyr, terminating with the buckling; (4) Although stellar bars remain strong during the plateau, their pattern speed and size stay nearly constant. This unusual behavior of stellar bars follows from highly reduced gravitational torques due to DM bar being aligned with the stellar bar. The orbital analysis shows that delayed buckling  results from slow evolution of stellar oscillations along bar major and vertical axes, thus postponing the action of the vertical 2:1 resonance which pumps the rotational energy into vertical motions;  (5) Peanut/boxy shaped bulges form at the beginning of the plateau and grow with time; (6) Finally, strong bars in spinning halos can avoid fast braking, resolving the long-standing discrepancy between observations and $N$-body simulations. This behavior of stellar bars along the $\uplambda$- and DM density-sequences reveals a wealth of stellar bar properties which require additional study.
\end{abstract}

\begin{keywords}
methods: numerical --- galaxies: bar --- galaxies: evolution --- galaxies: formation --- galaxies: kinematics and dynamics --- galaxies: structure
\end{keywords}



\section{Introduction}
\label{sec:intro}

Galaxies are believed to be nearly universally immersed in dark matter (DM) halos, which acquire their angular momentum during decoupling from the Hubble flow travel the cosmic web. Numerical studies have shown that the halo spin is low compared to the maximally allowed spin. Hence, halos can serve as an angular momentum sink for the embedded galactic discs, which spin at nearly the maximal rate \citep[e.g.,][]{sell80,deba00,atha02,marti06,villa09,villa10}. The strongest coupling between the discs and their host DM halos can be provided by stellar bars which torque the DM and represent the most significant departures from axial symmetry. 

However, the angular momentum transfer between a disc and its host DM halo is full of intricacies. Even nonrotating halos, i.e., with a spin\footnote{The DM halo spin, $\uplambda$, is defined as in \citet{bull01}, namely, $\uplambda= J/[\sqrt{2} M_{\rm h} r_{\rm h} v_{\rm c}(r_{\rm h})]$, where $J$ is the total angular momentum of the DM, and $M_{\rm h}$ and $r_{\rm h}$ are virial DM mass and radius, and $v_{\rm c}(r_{\rm h})$ is the circular velocity at $r_{\rm h}$.} $\uplambda = 0$  consist of prograde and retrograde angular momenta, cancelling each other. These prograde DM orbits, with respect to the embedded galactic discs, are coupled to disc orbits differently from the retrograde orbits, i.e., those that spin in the opposite direction \citep{coll19b}. The difference between the prograde and retrograde orbits is further amplified by the action of resonances. Naturally, in order to account for the resonant interactions in a disc-halo system, high resolution numerical simulations are required. Such resolution is easier to achieve in isolated galaxies than in cosmological models. New stellar dynamics effects are therefore better found, analyzed, and understood in isolated systems.

Bars have been known to be resilient morphological features which persist after being subjected to several kinds of strong perturbations \citep[e.g.,][]{sparke87}, but not all. For example, the merging of dense satellites \citep{pfenniger91} or destruction of their parent disc will naturally destroy the bars. Moreover, recent works on the evolution of stellar bars in spinning DM halos have questioned this paradigm, displaying how DM angular momentum can have a strong effect on stellar bar evolution \citep{coll18,coll19a,coll19b}. Differentially spinning DM halos increase the disc-halo interactions, amplify the rate of the angular momentum transfer, speed up the bar instability \citep{saha13,long14}, and, what is probably most intriguing, weaken and damp stellar bars \citep{coll18}. 

The reason why this phenomenon has avoided discovery for so long lies in that isolated galaxies have been modeled in rigid, or nonrotating halos, while cosmological simulations lacked resolution. \citet{coll18} models have analyzed the DM halo spin-sequence with $\uplambda = 0 - 0.09$, which included the vast majority of the parameter space. This model sequence encompasses essentially all spins encountered in numerical simulations of DM halos \citep[e.g.,][]{bull01}. 

\citet{coll18} halo models have identical masses and concentration. However, the additional parameter space of DM density appears to be unexplored. Yet, it is known that the bar dynamical friction against the host DM halo, i.e., the bar slowdown, depends on the background DM density \citep[e.g.,][]{deba00,atha03}. However, the importance of the halo spin in this respect has never been tested. 

In the present work, we widen this parameter space accounting for halos with different DM densities. Our results indeed show that varying this parameter leads to an interesting new effect, which we analyze here.   

Closely associated with the bar evolution, is the vertical buckling instability which develops after stellar bars reach their maximal strength. It has been argued that rotation plays no role in the buckling instability \citep[e.g.,][]{merr94}. That stellar bars are most suitable for buckling when strong, has been predicted theoretically \citep[e.g.,][]{toom66,kuls70,kuls71,mark71,araki85} and modeled numerically in pure stellar discs \citep[e.g.,][]{comb81,comb90,friedli90,raha91,pfen91,marti04,dubi09,villa09} and in two-component gaseous-stellar discs \citep[e.g.,][]{bere98,debat06,bere07,villa10}. A recurrent buckling of stellar bars has been also determined and analyzed in \citet[][see also review by \citet{shlo13}]{marti06}.

The physical nature of the buckling instability in stellar bars, however, is still under debate. Often in the literature this instability is not distinguished from the formation of peanut-shaped bulges in bars, yet \citet{friedli90} showed that peanut-shaped bulges can form even when suppressing buckling, so they are distinct processes: buckling leads generally to a peanut shape, but a peanut shape can occur without buckling. 

Two alternatives for explaining buckling or boxy/peanut shapes have been proposed, namely, the firehose instability \citep[e.g.,][]{raha91} and a resonance triggering\footnote{Here we follow the arguments given in \citet{li23}, which analyze the triggering factors for vertical buckling instability, namely, the firehose instability {\it versus} the asymmetric resonance triggering.} \citep[e.g.,][]{comb90}. 

\citet{li23} have determined that buckling in the nonrotating halos is associated with an abrupt increase in the central mass concentration in the bar, and triggers velocities along the bar and along the rotation axis, forming circulation cells which are absent in classical firehose instability. Analyzing the role of the vertical and planar 2:1 resonances, they found that the vertical 2:1 resonance appears only with the buckling and quickly reaches the overlapping phase, thus supporting the energy transfer from horizontal to vertical motions. Stars cross the vertical 2:1 resonance simultaneously with the buckling. They appear trapped in the planar 2:1 resonance at the same time, pointing to close relationship between the bending of stellar orbits and the resonant action.  

\citet{raha91} observed the weakening of stellar bars after the vertical buckling instability in nonrotating halos, and suggested their possible dissolution. The latter was never observed in numerical simulations, and is not expected, as shown by \citet{marti04}. On the other hand, spinning halos with $\uplambda > 0.03$ exert a two-fold effect on the bar evolution. The DM spin hastens the bar formation --- a dynamical stage in the bar evolution \citep{saha13,long14,coll18}. At the same time, the instability weakens the bar and essentially dissolves it leaving behind a weak oval distortion, as shown for spherical, oblate and prolate halos \citep{long14,coll18}. Note that \citet{peter16} have limited their halo spin to a single value of $\uplambda = 0.03$, and therefore did not investigate higher DM spins, where the increased feedback onto the bar evolution occurs.

The transfer of angular momentum from a barred disc to the DM halo leads to the trapping of DM particles by the stellar bars, forming the so-called "ghost" or "shadow" DM bar \citep{atha05,atha07,bere06,peter16,peter19,peter21,coll18,coll19a,coll19b}. The dissolution of the stellar bar leads to an immediate disappearance of the shadow bar \citep[][]{shlos08}.

Furthermore, the introduction of a nonzero DM spin amplifies the DM nonlinear response to the stellar bar by increasing the strength of the induced shadow DM bars by a factor of 3--4, when $\uplambda$ varies in the range of 0--0.09 \cite{coll19a}, who employed the orbital analysis to relate this effect to the DM orbit trapping byt the stellar bar. This is contrary to the conclusion by \citet{peter16} which claimed that the shadow bar strength is not affected appreciably by the DM rotation. Lastly, a counter-rotating halo with respect to the disc leads to another effect, the DM orbit spin reversal, and increased absorption of angular momentum by the counter-rotating DM halo \citep{coll19b}. 

Present work aims at analyzing the evolution of stellar bars along $\uplambda$ and DM density sequences, focusing on the angular momentum transfer between the {\it spinning} DM halos and their embedded barred stellar discs. We analyze both dynamical and secular evolution of stellar bars over a time period of 10\,Gyr, which includes the bar formation, its vertical buckling and formation of the boxy/peanut-shaped bulges. We provide a comparison with other works in section\,\ref{sec:discuss}.

This paper is structured as following. We discuss our numerical methods in Section\,2, and present the results of our numerical simulations in Section\,3. Section\,4 is devoted to discussing and analyzing our results, and we present our conclusions in the last section.

\section{Simulations and numerical methods}
\label{sec:numerics}

We use the $N$-body part of the mesh-free hydrodynamics code \texttt{GIZMO} \citep{hopk15}, an extension of \texttt{GADGET-2} code \citep{sprin05}. The models of stellar disc galaxies embedded in spherical DM halos have been constructed with different DM densities and a sequence of cosmological spins $\uplambda$. The units of mass, length, and velocity have been chosen as $10^{10}\,\mathrm{M_{\odot}}$, $1\,\mathrm{kpc}$, and $1\, \mathrm{km\,s^{-1}}$, respectively. As a result, the unit of time is approximately $1\,\mathrm{Gyr}$. The number of DM halo particles has been taken as $N_\mathrm{DM}=7.2\times 10^6$, and stellar disc particles $N_\mathrm{S}=8 \times 10^5$, in order to have a similar mass-per-particle. The gravitational softening length for both DM and stellar particles is $\epsilon_{\mathrm{DM}} = \epsilon_{\mathrm{S}} = 25 \, \mathrm{pc}$. The opening angle $\theta$ of the tree code has been reduced from $0.7$ used in cosmological simulations to 0.4 for a better quality of force calculation. And the tolerance of force accuracy has been set to 0.0001 to achieve a better conservation of energy and momentum. All models have been run for $10\,\mathrm{Gyr}$, with the angular momentum conservation within 0.2\% and energy conservation within 0.1\%.

To verify the numerical convergence, we have increased the precision calculation of the gravitational force by a factor of 10 and reproduced the standard model, RT180, evolution for $\uplambda=0.09$. We have also tested and found that the stellar disc does not thicken during the initial few 100\,Myr because of numerical heating. Finally, we have run test models with double gravitational softening and found no difference in the evolution either.

\subsection{Initial conditions}
\label{sec:ICs}  

The initial conditions for the stellar component consist of an exponential disc with a density profile as a function of a cylindrical radius $R$ given by
\begin{equation}
    \rho_{\mathrm{d}}(R, z) = \left( \frac{M_\mathrm{d}}{4 \pi R_0^2 z_0} \right) \, \exp \left(-\frac{R}{R_0} \right) \, \sech^2 \left( \frac{z}{z_0} \right),
    \label{eq:rho_d}
\end{equation}
where $M_\mathrm{d} = 6.3 \times 10^{10}\, \mathrm{M_{\odot}}$ is the mass of the disc, $R_0 = 2.85\, \mathrm{kpc}$ is the disc radial scalelength, and $z_0 = 0.6\, \mathrm{kpc}$ is the disc scaleheight. We truncate the disc at $6 R_{\mathrm{0}}\sim 17\, \mathrm{kpc}$, i.e., at 98\% of its mass.

The initial DM component consists of a spherical halo with the the density profile as a function of spherical radius $r$ described by \citet[][hereafter NFW]{NFW96},

\begin{equation}
    \rho_\mathrm{h} (r) = \frac{\rho_{\mathrm{s}} }{[(r + r_\mathrm{c}) / r_\mathrm{s}] (1 + r / r_\mathrm{s})^2} \, e^{- (r / r_\mathrm{t})^2},
\end{equation}
where $\rho_{\mathrm{s}}$ is a normalization parameter, $r_{\mathrm{c}} = 1.4 \, \mathrm{kpc}$ is the size of the flat density core, and $r_{\mathrm{s}} = 10 \, \mathrm{kpc}$ is a characteristic radius. To obtain the finite DM halo mass, we use a Gaussian cut-off radius $r_{\mathrm{t}}$.  

In order to study the effect of the halo density on DM halos within the $\uplambda$-sequence, we construct the NFW halos with the same halo and disc masses, by varying the Gaussian cutoff $r_\mathrm{t}$, thus producing the RT-sequences, as shown in Table\,\ref{tab:table1}. Hence, each RT sequence corresponds to the varying  concentration of the DM halo,
i.e., $r_{\rm h}/r_{\rm s}$.

\begin{table}
	\centering
	\caption{DM halo parameters for each RT sequence, where $\uplambda$ has been varied in the range 0--0.09 for prograde halos}
	\label{tab:table1}
	\begin{tabular}{lccc} 
		\hline
		Models & Halo Mass         & $r_{\rm t}$ & Average Halo Density         \\
		       & $10^{10}\,M_\odot$ & kpc         &   $ M_\odot\,{\rm kpc^{-3}}$ \\ 
		\hline
		RT86   & 63                & 86          & $2.4\times 10^5$             \\
		RT118  & 63                & 118         & $9.1\times 10^4$             \\ 
		RT150  & 63                & 150         & $4.4\times 10^4$             \\
		RT180  & 63                & 180         & $2.6\times 10^4$             \\
		RT210  & 63                & 210         & $1.6\times 10^4$             \\
		\hline
	\end{tabular}
\end{table}

\begin{figure}
\center 
	\includegraphics[width=\linewidth,height=\textheight,keepaspectratio]{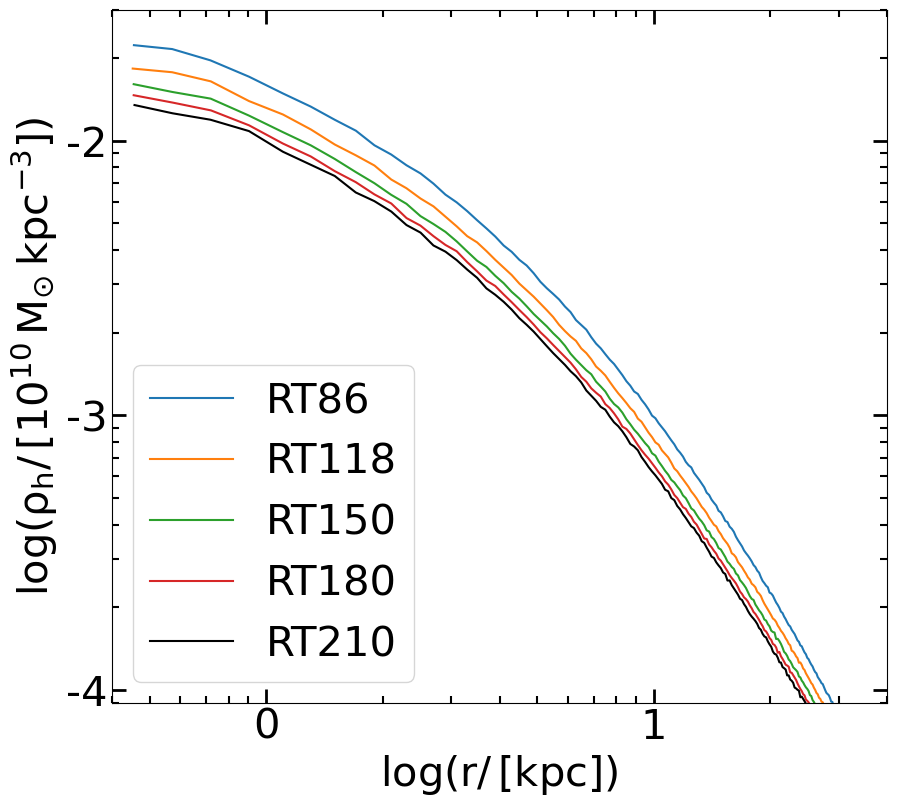}
    \caption{The halo density profiles along the RT-sequence at $t = 0$. These density profiles have been created by varying the Gaussian cutoff radius, $r_\mathrm{t}$, in the range of $r_{\rm t} = 86-210$\,kpc, and keeping the halo masses fixed (Table\,\ref{tab:table1}).}
    \label{fig:halo_rhoInit}
\end{figure}

We abbreviate the RT-sequence halos by specifying the value of $r_\mathrm{t}$ in kpc. We also abbreviate the $\uplambda$ value for each model by multiplying $\uplambda$ by 100. For example, the model with $r_\mathrm{t} = 180$\,kpc and $\uplambda = 0.045$ is named as RT180P45, where `P' stands for the prograde spinning halos with respect to the embedded discs. The initial halo density profiles for this RT-sequence are shown in Fig.~\ref{fig:halo_rhoInit}.

To assign the velocity distribution to each component, we use an iteration method introduced by \citet*{rodio06}, see also \citet{rodio09}, \citet{long14} and \citet{coll18}. For each iteration, we freeze the disc particles, then release the DM particles from the initial density distribution for $\sim 0.3\,\mathrm{Gyr}$. Next, we  return the DM particles to their nearest unevolved particles in the initial distribution, with the new velocities. Typically, we use $\sim 50-70$ iterations, until the virial ratio and the velocity distribution of DM particles become stable.

For the stellar disc, we use the epicycle approximation and asymmetric drift correction to determine the disc velocity. The stellar dispersion velocities are
\begin{equation}
    \sigma_{\mathrm{R}} (R) = \sigma_{\mathrm{R,0}} \, \exp \left( -\frac{R}{2R_0} \right)
\end{equation}
\begin{equation}
    \sigma_{\mathrm{z}} (R) = \sigma_{\mathrm{z,0}} \, \exp \left( -\frac{R}{2R_0} \right),
\end{equation}
where $\sigma_{\mathrm{z,0}} = 120\,\mathrm{km/s}$, and $\sigma_{\mathrm{R,0}}$ is determined by setting the minimal Toomre parameter \citep*[e.g.,][]{Binn08}, to $Q = 1.5$, at $\sim 2R_0$ in all models.

To spin up the halos, we randomly sample a fraction of retrograde DM particles (with respect to the disc rotation axis) and reverse the azimuthal velocities, $v_{\phi}$, of these particles to achieve the target $\uplambda$, in the range of 0--0.09. Changing the sign of $v_{\phi}$ does not affect the solution of the collisionless Boltzmann equation \citep{lynd62}.

\subsection{Evolution of the bar amplitude}
\label{sec:amplitude}

\begin{figure*}
\center 
	\includegraphics[width=0.7\textwidth]{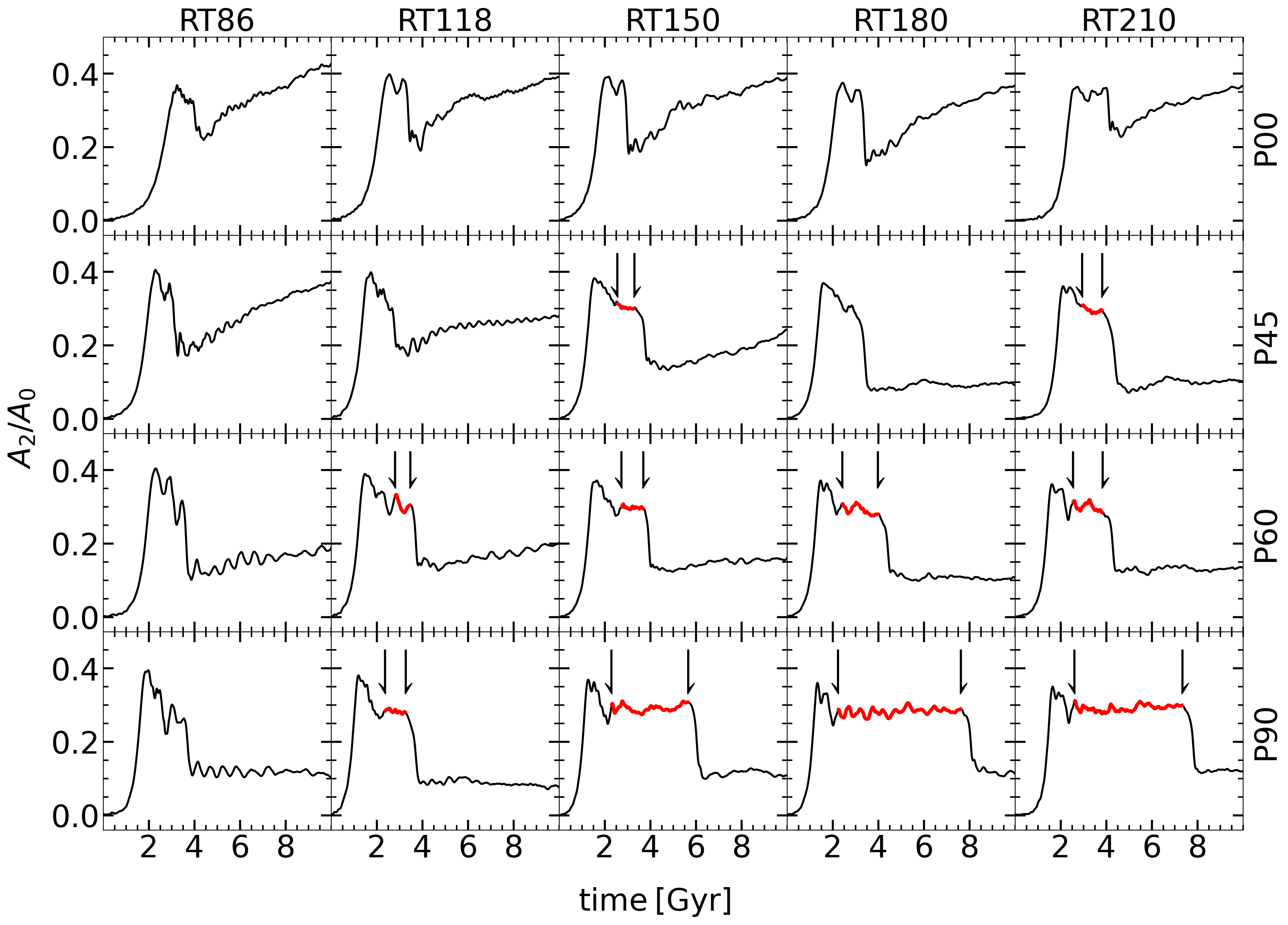}
    \caption{Evolution of normalized Fourier amplitudes, $A_2/A_0$, of stellar bars along the RT-sequence (horizontal rows). Each horizontal row displays models with the same $\mathrm{\uplambda}$ along varying $r_{\rm t}$ parameter. Vertical columns represent the $\mathrm{\uplambda}$ sequence, from P00 (top row), to P45, P60 and P90, respectively. The plateau stage has been emphasized by the red thick line between the vertical arrows.}
    \label{fig:a2RT}
\end{figure*}

In all models, stellar bars develop from the initially axisymmetric mass distribution. We quantify the bar strength by the Fourier components of the surface density. For $m$ mode, the Fourier amplitude is $\sqrt {a_{\rm m}^2 + b_{\rm m}^2}$, where 
\begin{equation}
    a_{\rm m}(r) = \frac{1}{\pi} \int_{0}^{2\pi} \Sigma(r, \theta) \, \cos(m\theta) \, \mathrm{d} \theta, \; \; \; \; \; m=0, 1, 2, \ldots ,
\end{equation}
\begin{equation}
    b_{\rm m}(r) = \frac{1}{\pi} \int_{0}^{2\pi} \Sigma(r, \theta) \, \sin(m\theta) \, \mathrm{d} \theta, \; \; \; \; \; m=1, 2, \ldots ,
\end{equation}
and $\Sigma(r, \theta)$ is the surface stellar density. To quantify the bar strength, we use the normalized $A_2$ amplitude which is defined as 
\begin{equation}
    \frac{A_2}{A_0}=\frac{ \int_{0}^{R_{\mathrm{max}}} \sqrt {a_{2}^2(r) + b_{2}^2(r)} \, \mathrm{d} r}{ \int_{0}^{R_{\mathrm{max}}} a_0(r) \, dr }.
    \label{eq:barA2}
\end{equation}
We choose the upper limit of integration, $R_{\mathrm{max}}$, as the radius which contains $98\%$ of the disc mass at a given time.  

Similarly, we measure the vertical buckling strength, $A_{1z}$, i.e. the vertical asymmetry, by calculating the the $m=1$ Fourier amplitude in the $xz$-plane, where the major axis of the bar is aligned with the $x$-axis, and the rotation axis is along $z$-axis, as follows,
\begin{equation}
    \frac{A_{1z}}{A_0}=\frac{ \int_{-x_0}^{x_0} \sqrt {a_{1}^2 + b_{1}^2} \, \mathrm{d} x}{ \int_{-x_0}^{x_0}a_0 \, \mathrm{d} x }.
    \label{eq:barA1z}
\end{equation}
The integral is over the region $|x| < 12$ kpc, $|y| < 3$ kpc, $|z| < 5$ kpc.

Lastly, the phase of the bar, $\phi_{\rm bar}$, is obtained from
\begin{equation}
\phi_{\rm bar} = \frac{1}{2} \arctan\left(\frac{b_2}{a_2}\right).    
\end{equation}
Generally, $\phi_{\rm bar}$ displays small variations, because of the noise in $b_2$ and $a_2$. We take an average $\phi_{\rm bar}$ in a range of $r$ which defines the bar size.

\subsection{Evolution of the bar size}
\label{sec:barsize}

We measure the bar size evolution using the orbital analysis method which is the most reliable \citep{hell96,bere98,marti06,coll18}. We start by computing the fundamental orbit family, the $\mathrm{x_1}$ family in the notation of \citet{cont80}, which constitutes the backbone of a stellar bar. Next, we determine the extent of this family to the highest value of Jacobi energy. This family is terminated inside the CR. 

\section{Results}
\label{sec:results}

We present two model sequences here, namely, the RT-sequence of decreasing DM halo density, by increasing the halo's $r_{\rm t}$ from 86\,kpc to 210\,kpc (see Table\,\ref{tab:table1}), and the $\uplambda$-sequence. For each halo size, we produce four models of spin $\uplambda=0$,  0.045, 0.06 and 0.09 (section \ref{sec:resultsRT}). The $\uplambda = 0.045$ sequence is chosen as a transition sequence between models that show the effect of the spin and those that do not show. To carefully assess the effects of the halo rotation, we choose the standard RT180 model, where the spin is varied in smaller increments, $\uplambda = 0$, 0.03, 0.045, 0.06 and 0.09 (section \ref{sec:resultsRT180}). This RT180 model displays the new effect clearly, unlike the transition model sequence with $\uplambda = 0.045$.

\subsection{RT model sequence: changing the DM halo density}
\label{sec:resultsRT}

We start by comparing the bar strength evolution along the RT sequence, displaying also the $\uplambda$ dependence of these models (Figure\,\ref{fig:a2RT}). The P00 sequence shows a very weak and basically non-existent dependence on the halo density. For this $\uplambda$, the strong decline of $A_2$ after the initial growth corresponds to the vertical buckling instability in the bar, as we have verified. 

This dependence on the DM density, however, shows up in the transition RT models of P45 in two different ways. First, the bars have difficulty to regain their strength after buckling --- already RT118 shows a clear saturation in the bar strength during its secular evolution. And for models RT180 and RT210, the bars are essentially dissolved after buckling. 

Second, we observe a marginal appearance of a plateau in $A_2$ starting with the RT150P45 model. This plateau reflects the postponement of buckling by $\Delta t\sim 0.7$\,Gyr. The length of this plateau increases gradually following the RT and $\uplambda$ sequences, as we discuss below. We have verified that the initial decline in $A_2$ by $\sim 15\%-20\%$ in all models does not result in  the breaking of the vertical symmetry in the stellar bars. The actual break comes after the plateau stage.

\begin{figure}
\center 
	\includegraphics[width=0.48\textwidth]{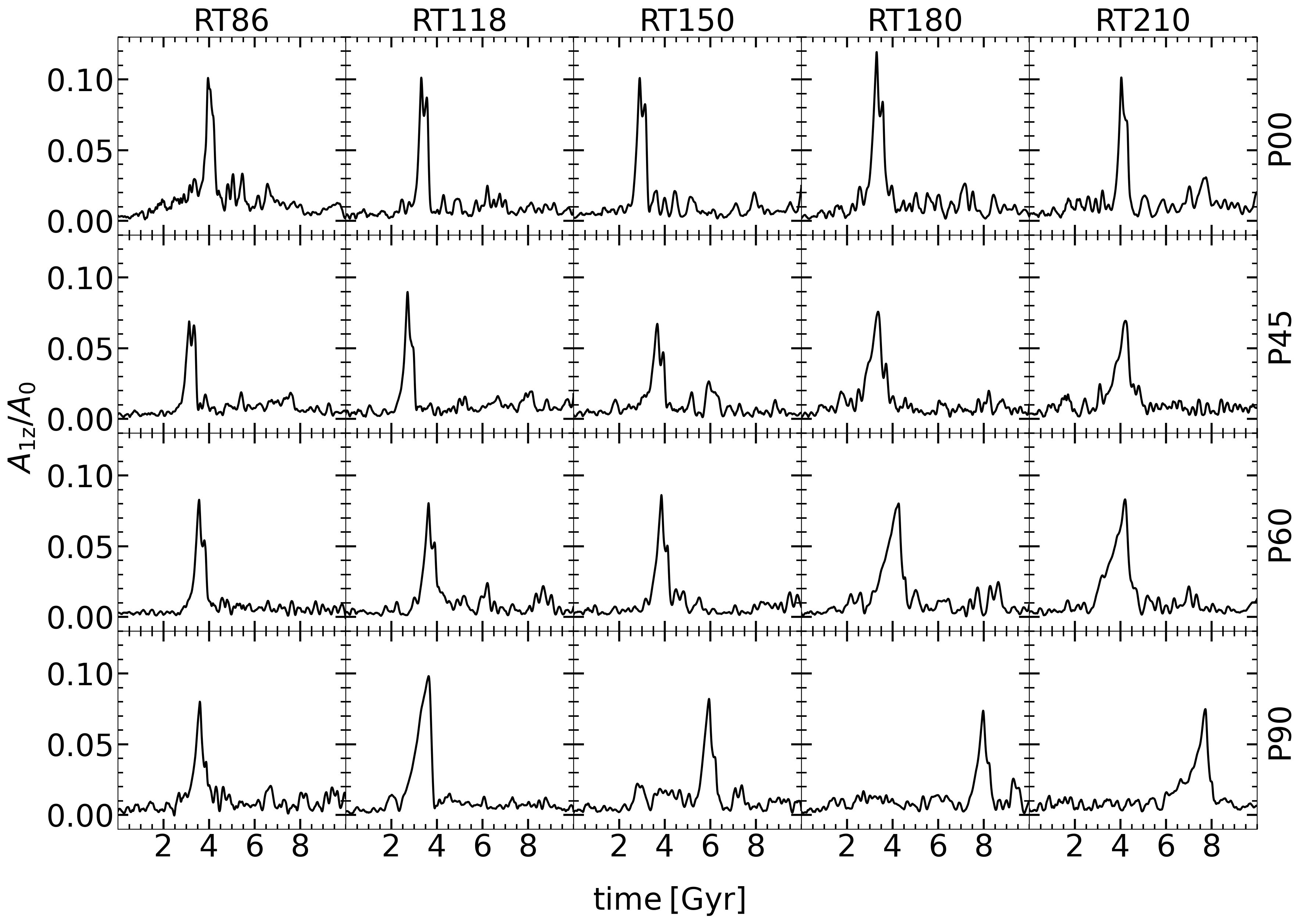}
    \caption{Buckling amplitude of the vertical $m=1$ mode of stellar bars, $A_{1z}$, normalized by $A_0$, along the RT and $\uplambda$ sequences. Note that the vertical bucklings happen only at the end of the plateau stage. }
    \label{fig:A1z_rt-all}
\end{figure}

\begin{figure}
\center 
    \includegraphics[width=0.48\textwidth]{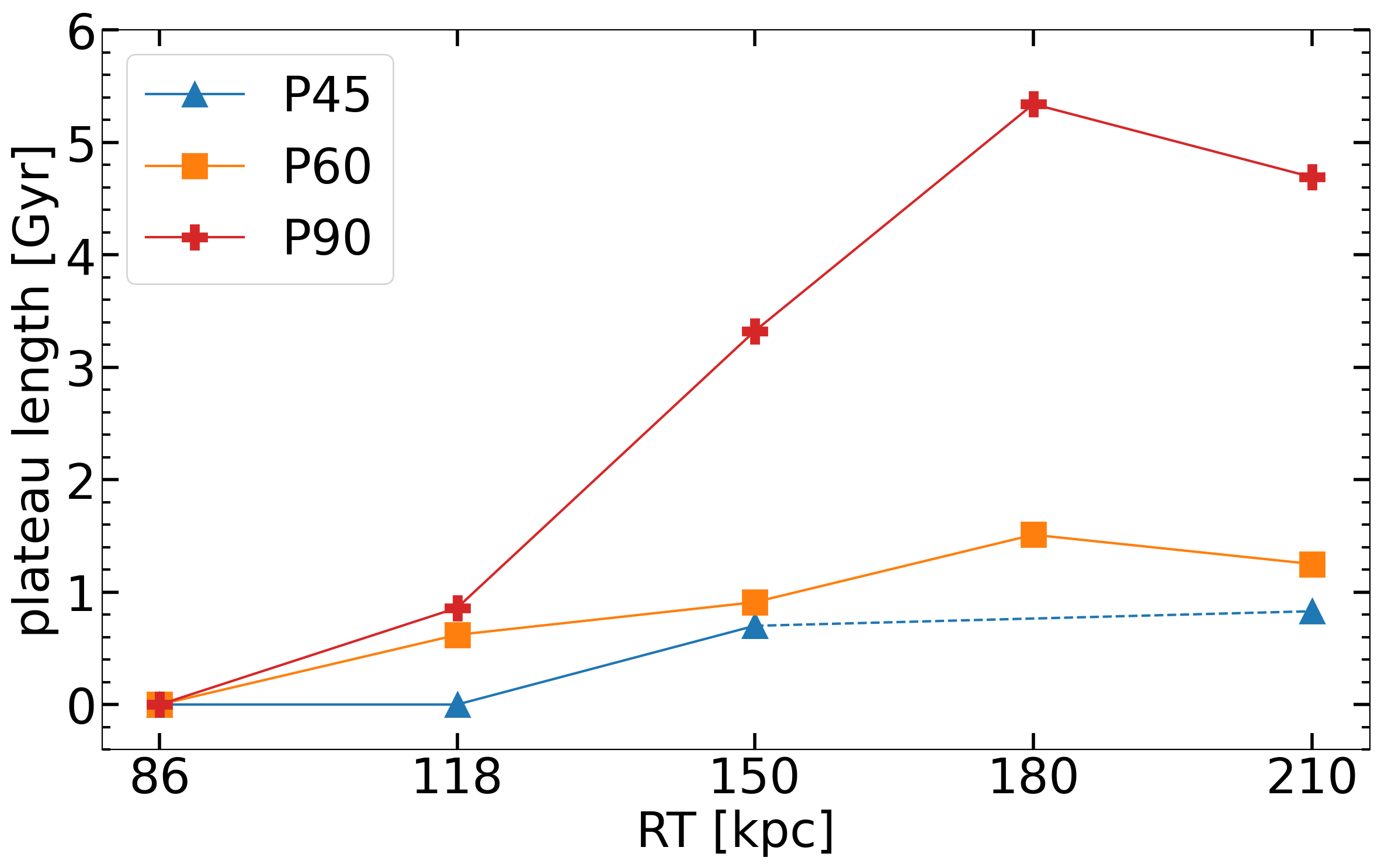}
    \caption{The length of the plateau in $A_2$ for models that exhibit it. The plateau in RT180P45 model being ignored due to its short duration, less than 0.5\,Gyr. Due to the difficulty to measure the plateau length for the RT180P45 model, its corresponding point on this diagram has been omitted. See the text for relevant details in measuring the plateau length.}
    \label{fig:plateau_length}
\end{figure}

\begin{figure*}
\center 
	\includegraphics[width=0.7\textwidth]{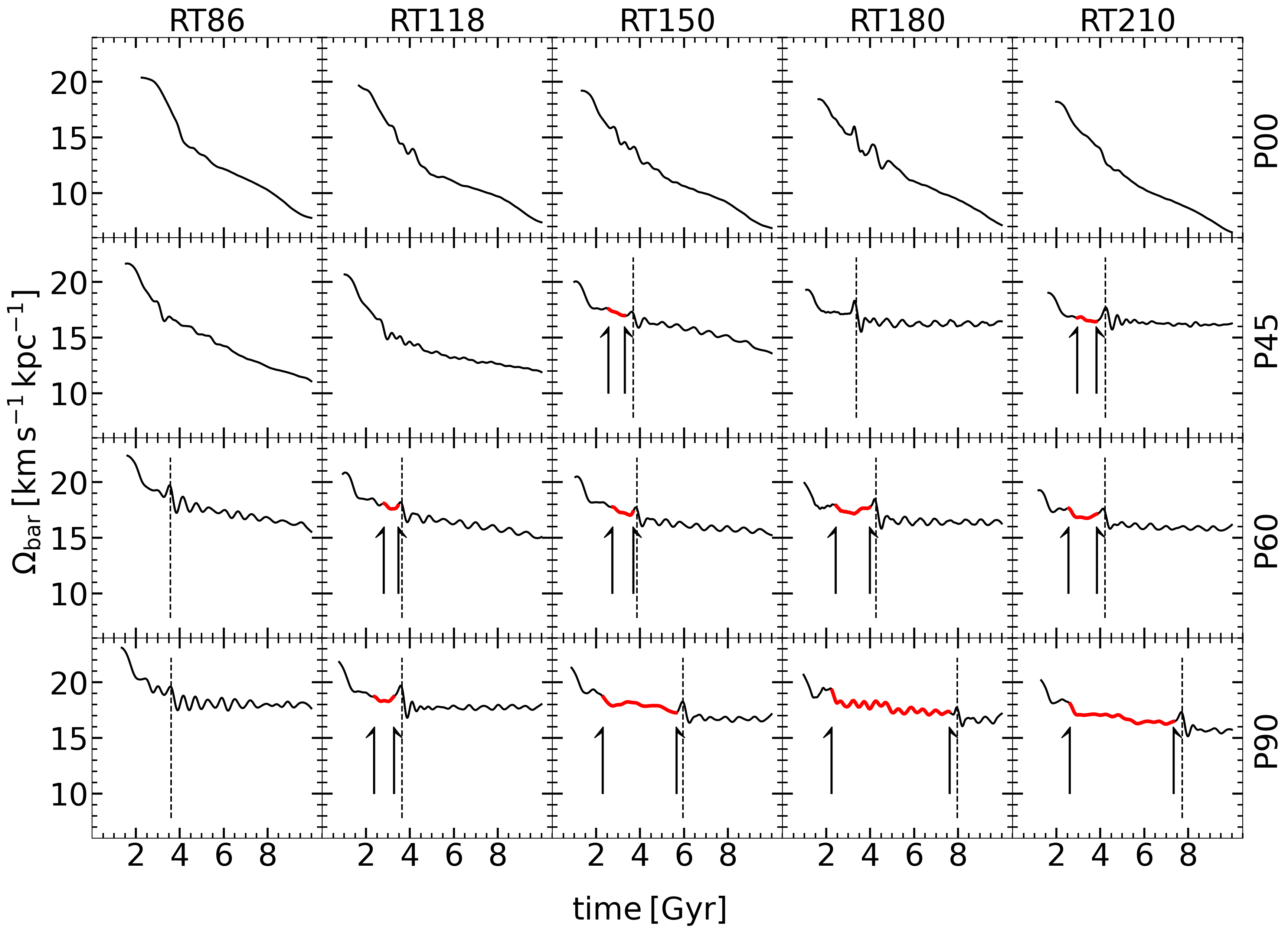}
    \caption{Stellar bar pattern speeds, $\Omega_{\rm bar}$, evolution along the RT and $\uplambda$ sequences. The plateau stages are colored red, and are found between the vertical arrows. The vertical dashed lines indicate the time of the maximum vertical buckling in some models. After the buckling, the stellar bars in these models dissolve and form oval distortions.}
\label{fig:omgBar-all}
\end{figure*}

\begin{figure}
\center 
	\includegraphics[width=0.48\textwidth]{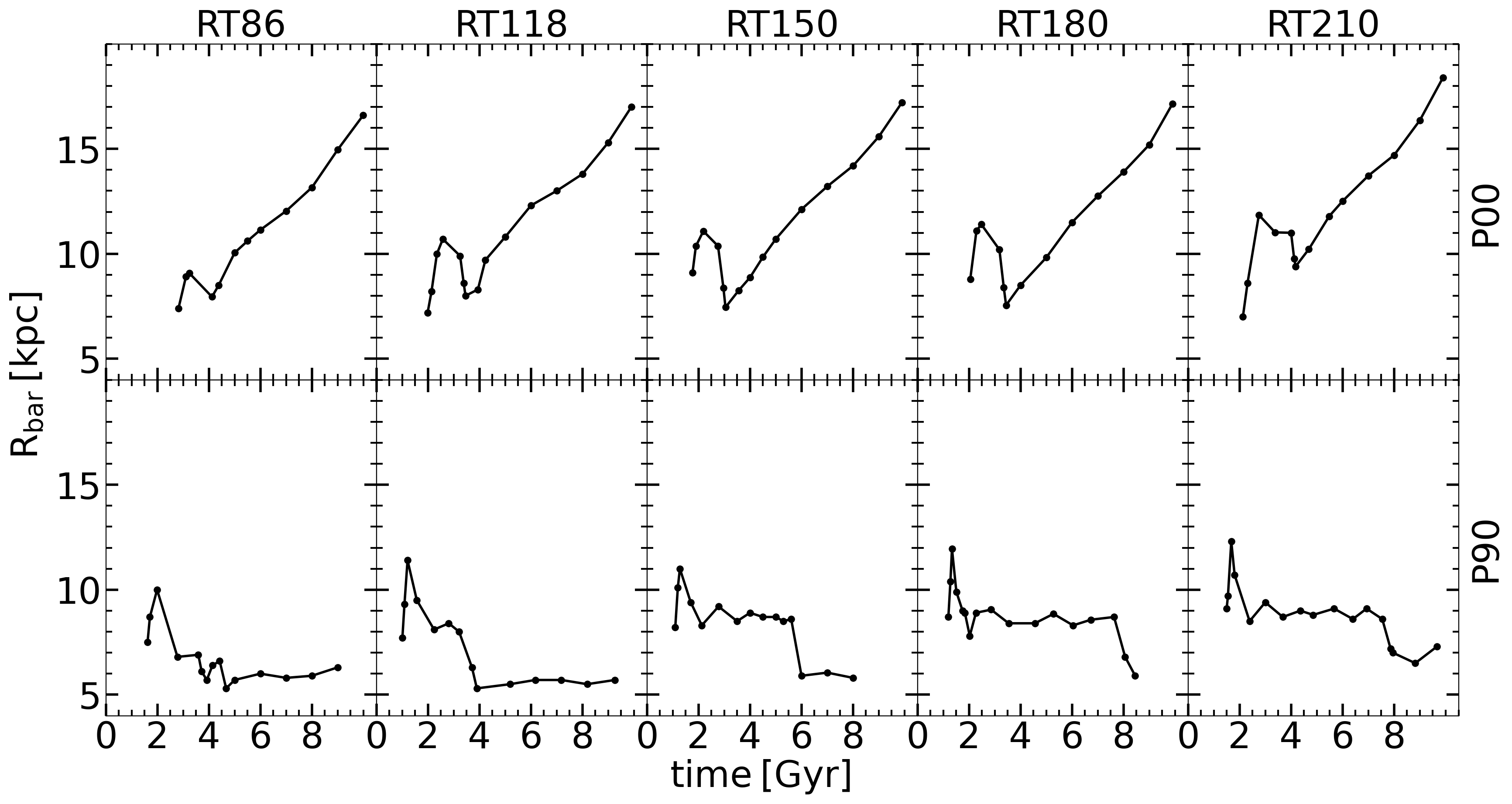}
    \caption{Bar size evolution along the RT sequence, for P00 models (top) and P90 models (bottom). Note that the initial evolution is similar in all models, until the $A_2$ reaches its maximum. Subsequently, the bars shorten, and, while the $\uplambda=0$ bars resume their growth, the $\uplambda=0.09$ bars remain of the same size until the end of the plateau period. Thereafter they buckle and dissolve, leaving behind a weak oval distortion.}
    \label{fig:barSize}
\end{figure}
 
\begin{figure*}
\center 
	\includegraphics[width=0.7\textwidth]{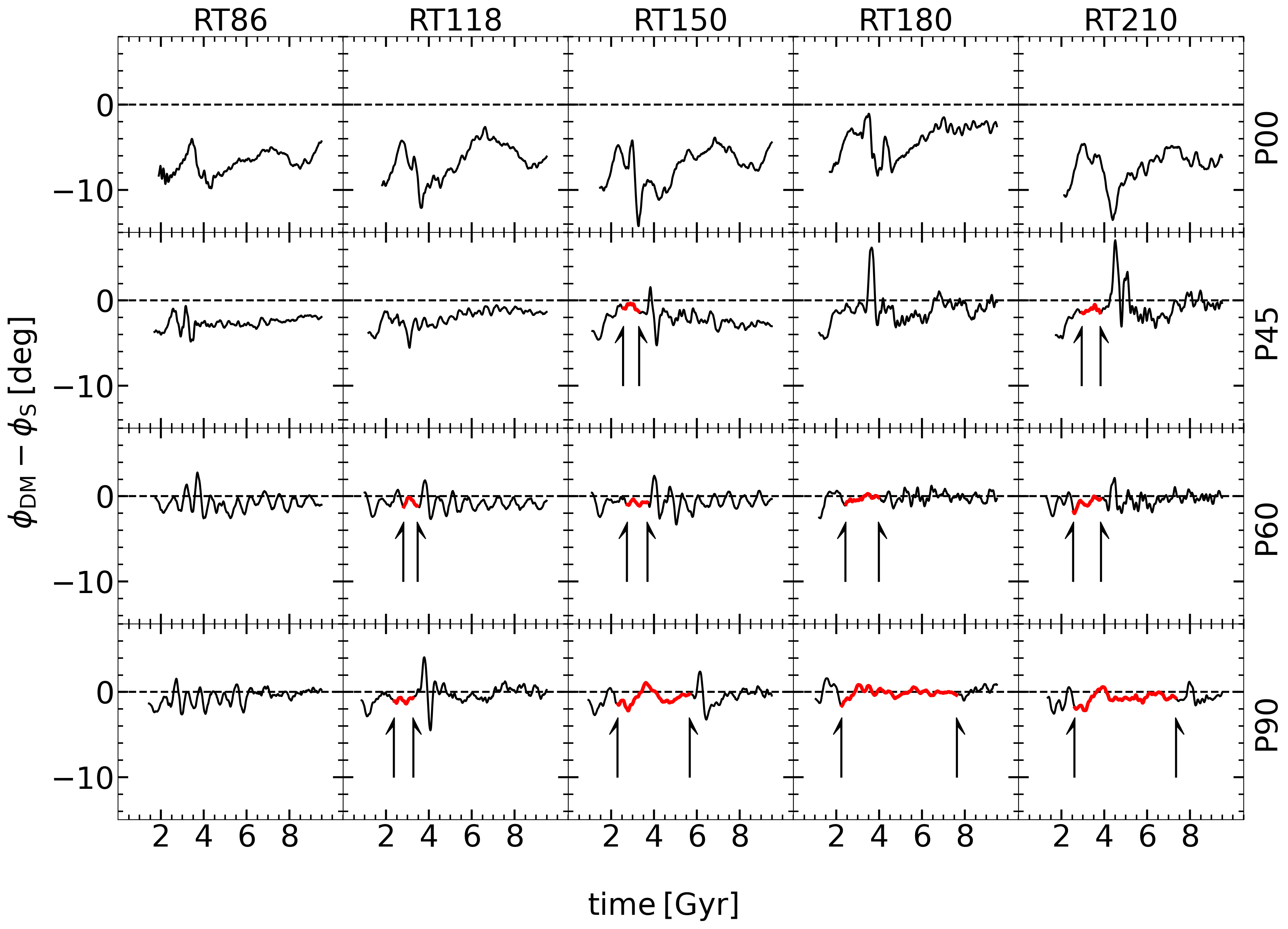}
    \caption{Angle evolution between the DM and stellar bars along the RT and $\uplambda$ sequences, $\Delta\phi = \phi_{\rm DM} - \phi_{\rm S}$. This angle is measured from the DM bar to the stellar bar and hence is either negative or zero. The plateau stages are colored in red and situated between two vertical arrows. Note the time of buckling is characterized by larger amplitude wiggles. The horizontal line lies at  $\Delta\phi = 0$ for reference. The angle $\Delta\phi$ is steadily decreasing along the RT and along the $\uplambda$ sequences, and is basically zero for all P60 and P90 models.}
    \label{fig:dphi_h2d_rt-all}
\end{figure*} 

We note that the choice of four $\uplambda$ sequences shown in Figure\,\ref{fig:a2RT} and subsequent Figures is made on the basis of many more sequences analyzed in \citet{coll18}. Among these sequences, models with $\uplambda \ltorder 0.04$, display no effect of spinning halos on the stellar bar evolution. While models with $\uplambda > 0.04$ do show progressively stronger effect with increasing $\uplambda$.

The P60 sequence displays the plateau phase in all models except the RT86 one. The duration of the plateau increases and doubles from 0.9\,Gyr in RT118 to 1.8\,Gyr in RT210. The buckling of the stellar bar is delayed by the plateau duration and happens at its end. Starting with model RT150, the bar does not regrow after buckling, leaving behind an oval distortion.

Evolution along the P90 sequence is the most interesting and dramatic one. In the RT86, $A_2$ declines to $\sim 0.1$ after buckling and levels off thereafter --- the bar never recovers and only a weak oval distortion remains. But in RT118 and the rest of the sequence, $A_2$ declines by about 20\% after reaching its maximum, without buckling. Then it levels off, forming a plateau. The time duration of this plateau increases with $\uplambda$, from $\sim 1$\,Gyr in RT118 to $\sim  5$\,Gyr in RT210. Note that the bar strength during the plateau stage is the same in all the models. The plateau ends in buckling and in subsequent bar dissolution in these models. 

We have tested the initial decline in $A_2$, preceding the plateau, and find that it is not associated with the vertical buckling, as shown by the vertical $m=1$ mode amplitude, $A_{1z}$ (Fig.\,\ref{fig:A1z_rt-all}). Nevertheless, we do observe an appearance of a peanut/boxy bulge shape during the plateau stage. Along the plateau, the peanut/boxy bulge strengthens and grows. 

To further analyze the basic properties of stellar bars in our models, we display the bar pattern speeds evolution, $\Omega_{\rm bar}$, in Figure\,\ref{fig:omgBar-all}. The top row exhibits the classical slowdown of stellar bars in nonrotating halos. For the $\uplambda=0$ sequence, $\Omega_{\rm bar}$ slows down by a factor of $\sim 2-3$ {\it after} the buckling, i.e, over a time period of $\sim$ 6\,Gyr. For $\uplambda=0.045$ sequence, the slowdown is by a factor of $\sim 0-2$, for $\uplambda=0.06$ sequence, it slows down by a factor of $\sim 0.1-0.15$, and for $\uplambda=0.09$ the slowdown is a factor of only $\sim 0-0.1$.
 
For spinning halos, we focus on $\Omega_{\rm bar}$ during the plateau time. The pattern speeds appear nearly constant for these periods of time, which is clearly surprising, as all these bars are strong and should experience braking against the DM halo due to the associated angular momentum transfer. This effect is especially pronounced in the P90 models, where the bar does not experience braking over very long periods of time, of a few Gyr. Consequently, the time periods of $\Omega_{\rm bar}\sim \mathrm{const}$ correspond to a fixed CR radius.

We note that the actual buckling of stellar bars happens only after the plateau stage, as shown in Figure\,\ref{fig:A1z_rt-all}. By buckling event, we mean that the bar breaks its vertical symmetry, forms a boxy/peanut bulge\footnote{Note that the boxy/peanut bulges can form under alternative conditions, as we discuss below.}, and dramatically shortens its size. However, the first slight reduction in $A_2$ after reaching its maximal value only leads to a small contraction in the bar length, as evident from Figure\,\ref{fig:barSize}, and to no breaking of its vertical symmetry. We, therefore, do not consider it being a buckling. Note that the stellar bar size does not grow during the plateau stage, which is consistent with the stagnation of the CR radius (Fig.\,\ref{fig:barSize}).

The `plateau' in $A_2$ refers to a steady state that begins after a $\sim 20 \%$ decline of $A_2$ curve from its global maximum and ends at the vertical buckling (which is followed by a substantial decline in $A_2$, as shown in Figure\,\ref{fig:a2RT}). In addition, during the plateau stage, the slowdown rate of the $\Omega_{\mathrm {bar}}$ is relatively small compared to that of the standard P00 models. 

We have used the modified Ramer–Douglas–Peucker algorithm \citep{doug73} in order to determine the plateau length in $A_2$ (Fig.\,\ref{fig:plateau_length}). The endpoint of the plateau is taken as the buckling time. Next, we proceed from this point towards the $A_2$ global maximum, i.e., to the left, by finding the turning points on the $A_2$ curve which correspond to the local maxima, and connecting the left maximum with the endpoint. We measure each peak amplitude defined as $\left| \Delta A_2 \right|$, between the endpoint and each peak before the endpoint. The starting point of the plateau is defined as the first peak behind which (i.e., to the left)  the parameter $\left| \Delta A_2 \right|$ exceeds the critical value, i.e., $\left| \Delta A_2 \right| > 0.03$.  

Additionally, we set a lower limit of 0.5\,Gyr to identify the minimal length of the plateau, to avoid confusion. For example, the plateau in the RT180P45 model is too short and is ignored. On the other hand, in RT210P00, the short flatness region in $A_2$ has been excluded because its $A_2$ does not feature the $\sim 20 \%$ decline from the global maximal maximum before the buckling. Moreover, the slowdown of $\Omega_{\mathrm {bar}}$ in this model does decrease sharply, and hence this is not a steady state.

The simplest and probably the only explanation for the absence of the bar slowdown during the plateau stage is that the stellar bar does not experience an external torque, or that the sum over all the existent torques vanishes. The only torques that can be applied are those from the outer disc (i.e., outside the CR), from the nonrotating or slowly spinning bulge, or from the gravitational wake in the DM triggered by the tumbling stellar bar, i.e., from the DM bar. The first option is possible as long as the CR lies inside the disc. It typically acts during the bar formation itself, and weakens substantially thereafter, because, as the CR moves out, there is very little mass in the outer disc which is capable of absorbing the angular momentum from the bar. So at best, it is a very weak recipient of the angular momentum. The second option is absent in the RT sequence models which have been constructed without the classical bulge. On the other hand, the third option remains viable at all times.

As the DM responds to the stellar bar torquing by forming DM wakes, we perform the Fourier analysis for the DM mass distribution surrounding the stellar disc, and find the position angle of the induced DM $m=2$ mode. This is performed in the following way in order to avoid the spiral arms in the stellar disk and the spiral wakes in the DM halo --- both induced by the stellar and DM bars, respectively. During the Fourier analysis we require the phase angle of the $A_2(r)$ to be constant. Hence, our stellar and DM bar lengths do not include the associated spirals and wakes.

Figure\,\ref{fig:dphi_h2d_rt-all} exhibits the angle between the DM to stellar $m=2$ modes as a function of time. This angle is expected to be always negative because the stellar bar triggers and is always leading the DM response\footnote{During the time of buckling, the measurement of the position angles of DM and stellar bars from the $m=2$ mode is inaccurate due to the misalignment of the $m=2$ component of the surface density contours, and results in unreliable angle between the DM and stellar bars.}.  All nonspinning halos show that the stellar bar is indeed leading the DM bar by about $10^\circ$. This means that a substantial torque is applied on the stellar bars, resulting in efficient braking.

But models with $\uplambda = 0.045$ display a much smaller angle by a factor of 2-3. For RT180 and RT210, the angle is very close to zero, as in all the models of the P60 sequence, and this results in a very small braking\footnote{Note that we only discuss here the angle between the bars before the buckling. After the buckling, the stellar bar essentially dissolves and the DM wakes vanish accordingly.}. 

Finally, models with $\uplambda = 0.09$ exhibit DM and stellar bars being essentially aligned, except during the buckling, when it is conceptually difficult to calculate the position angle of the bar. These stellar bars display a remarkably steady tumbling for periods reaching $\sim 7$\,Gyrs.

As long as this angle is non-zero, the stellar bar will experience gravitational torque from the DM bar and brake. In essence, this is the dominant reason for the angular momentum transfer from the stellar disc to the DM halo in our models, subject to resonance and nonresonant couplings of DM and stellar bars.

We use Figure\,\ref{fig:a2RT} in order to determine the timing of the plateau periods in the RT sequence, and correlate them with the angle between the DM and stellar bars in Figure\,\ref{fig:dphi_h2d_rt-all}. These plateau directions are relatively short in the P45 models, starting with RT150 and up to RT210, and the exact duration of the plateau is difficult to assess. But in P60 models, the plateau clearly stands out, and this difficulty does not exist. The plateau duration is also increasing along the RT sequence. In P90 models, the plateau is long lived and its duration shows the same trend as in P45 and P60.

Note that the braking disappears in two cases --- one, which is trivial, when the buckling dissolves the stellar bar. The second one, during the plateau, when the stellar and DM bars can be considered as strong bars. Note also that the DM bars appear to be $\sim 3-4$ times stronger and more massive in P90 models compared to P00 models, and the ratio of the stellar-to-DM in the bars increases with $\uplambda$ \citep{coll19a}.  

In summary, we observe a specific stage in the evolution of stellar bars in {\it spinning} DM halos with gradually reduced DM density, characterized by a relatively strong bar which nevertheless experiences a dramatically reduced braking against the DM, and displays evolution in its size. Models with a low spin, $\uplambda \ltorder 0.04$, and higher DM density do not display this behavior. This includes the nonrotating halos with $\uplambda=0$. We return to this issue in the Discussion section.

\subsection{The angular momentum transfer}
\label{sec:resultsRTJdot}

\begin{figure*}
\center 
	\includegraphics[width=0.8\textwidth]{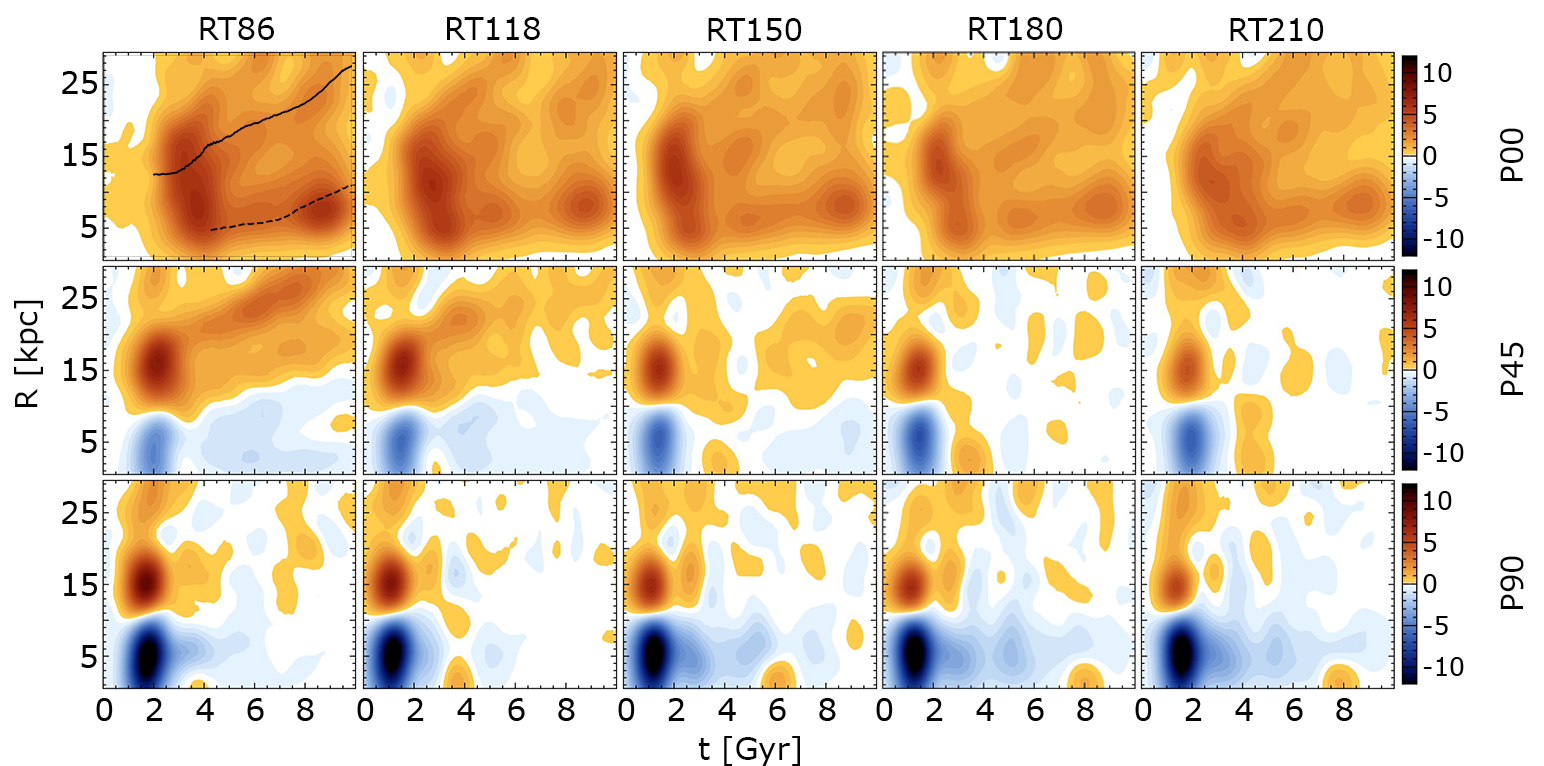}
    \caption{The rate of the angular momentum, $\dot{J}$, in the \emph{halo} along the RT-sequence, as a function of the cylindrical radius and time for P00 (top row), P45 (middle row), and P90 (bottom row) halos. $\dot{J}$ is calculated by every 0.05 Gyr, within a cylindrical shell $\Delta R = 1 \, \mathrm{kpc}$ and $|z| < 6 \, \mathrm{kpc}$. The color palette corresponds to gain/loss (i.e. red/blue) of the angular momentum. In P90 halos (bottom), the plateau end at the time where a small yellow area (the absorption of $J$) appears inside the $R \sim 5 \, \mathrm{kpc}$. The ILR radius (dashed line) and the CR radius (solid line) are plotted in the panel of RT86P00 model.}
    \label{fig:jdot_rtseq_halo}
\end{figure*}

\begin{figure*}
\center 
	\includegraphics[width=0.8\textwidth]{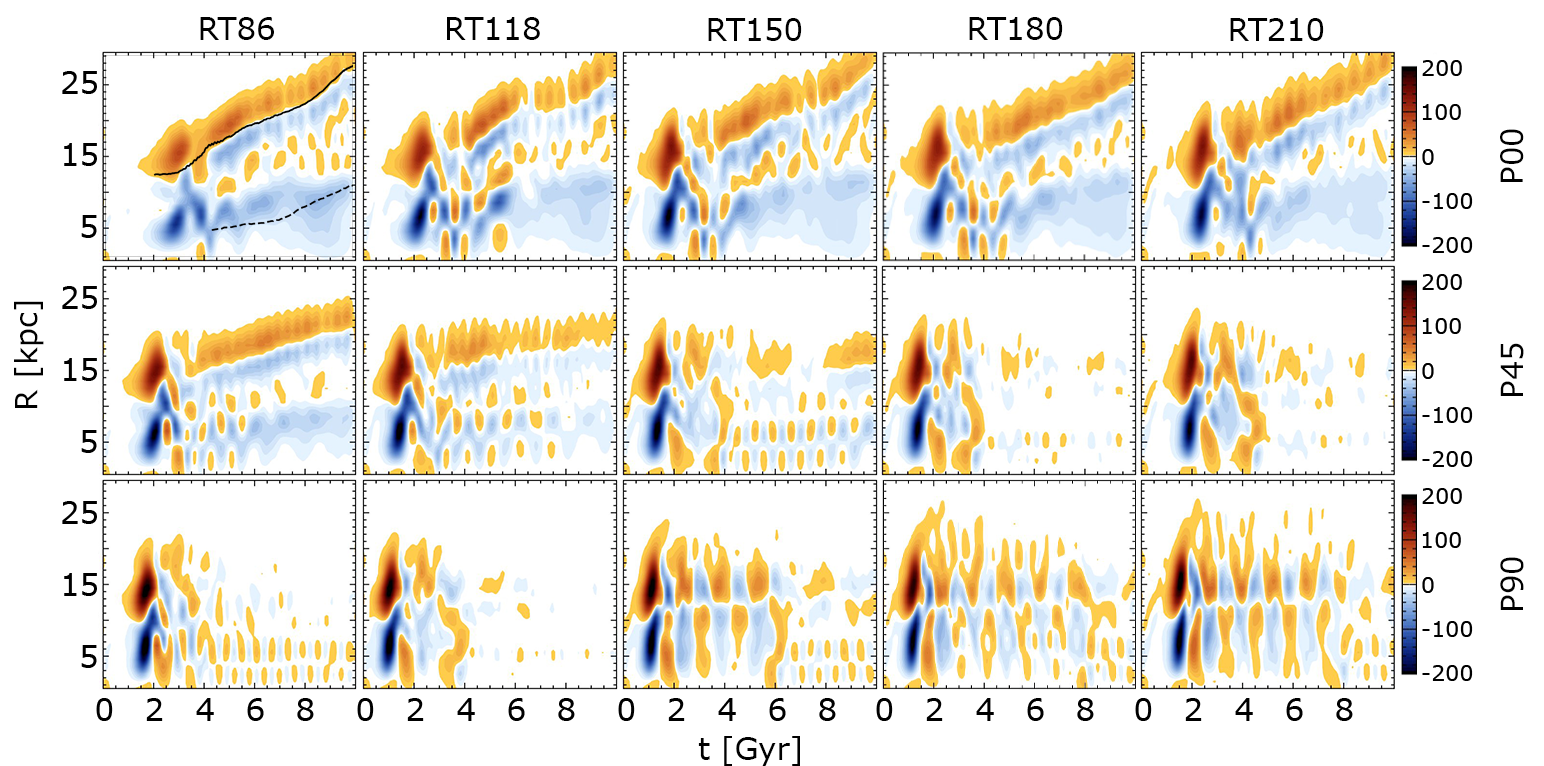}
    \caption{Same as Figure\,\ref{fig:jdot_rtseq_halo} but for the \emph{disc} embedded in  P00 (top row), P45 (middle row), and P90 (bottom row) halos along the RT-sequence. In P90 halos (bottom), the plateau stages basically correspond to the small oscillations of the gain and loss of $J$ around the CR $\sim 12\,\mathrm{kpc}$. Note the color bar range is different from Figure\,\ref{fig:jdot_rtseq_halo}.}
    \label{fig:jdot_rtseq_disc}
\end{figure*}

The angular momentum redistribution in disc-halo systems is the basis for evolution of stellar bars. To quantify the angular momentum transfer between the inner and outer disc across the CR resonance, as well as between the disc and the parent DM halo, we use the method introduced by \citet{villa09}, see also \citet{coll18}. For this purpose, we divide the discs and the halos into cylindrical shells in $R$ and calculate the rate of angular momentum change, $\dot J$, in each shell, which is integrated vertically for $|z| < 6\,\mathrm{kpc}$. The resulting evolution of $\dot J$ is then followed in time. Figures\,\ref{fig:jdot_rtseq_halo} and \ref{fig:jdot_rtseq_disc}  display the evolution of $\dot J$ in all models along the RT- and $\uplambda$ sequences, for DM halos and discs, respectively. 

The RT halo sequence with $\uplambda=0$ displays the DM halo, which exhibits absorption only of the angular momentum. The three main resonances \cite[e.g.,][]{Binn08}, the inner Lindblad resonance, ILR, the corotation resonance, CR, and the outer Lindblad resonance, OLR,  stand out, having elevated $\dot J$. The bar slows down and the resonances move out. The bar strength correlates nicely with $\dot J$. 

The RT halo sequence with $\uplambda=0.045$ appears to differ substantially from $\uplambda=0$ in Figure\,\ref{fig:jdot_rtseq_halo}. Most importantly, we observe that the halo emits angular momentum as well as absorbs it. The emission of $\dot J$ culminates at the time of the buckling in all models, but its strength increases along the RT sequence as well. This emission is associated with the halo ILR, as evident from Figure\,\ref{fig:jdot_rtseq_halo}, see also \citet{coll19a}. Absorption of the angular momentum, on the other hand, happens at CR and OLR. It weakens with the RT sequence, as it is correlated with the bar strength, which does not grow after buckling or dissolves. This emission and absorption of angular momentum by the halo has its origin in the formation of the DM bars in spinning halos --- these bars correspond to the gravitational response of the DM to the stellar bars. The inner halo close to the disc is losing its angular momentum to the DM lying further out.

The last row of Figure\,\ref{fig:jdot_rtseq_halo} has $\uplambda=0.09$ and is in sharp contrast with $\uplambda=0$ sequence. Emission of halo angular momentum prevails over its absorption. After buckling, the stellar bar has dissolved and the transfer of angular momentum to the halo has ceased. Note that during buckling in $\uplambda=0.09$ models, the halo experiences absorption of the angular momentum for $r\ltorder 5$\,kpc. So, the buckling process has an effect on the $J$-redistribution in the disc-halo system, and symbolizes a fully 3-D process, as discussed in Li et al. (2022).

The RT disc sequence correlates well with the parent halo properties. The top row in Figure\,\ref{fig:jdot_rtseq_disc} shows that the outer disc does absorb some angular momentum from the bar region. The ILR always emits $J$. The angular momentum redistribution dies out after buckling as the bar weakens or dissolves. Interestingly, during the plateau stages in Figure\,\ref{fig:a2RT}, the angular momentum transfer between the halo and the disc is basically non-existent.

\subsection{RT180 model sequence: focus on DM spin}
\label{sec:resultsRT180}

\begin{figure*}
\center 
	\includegraphics[width=0.7\textwidth]{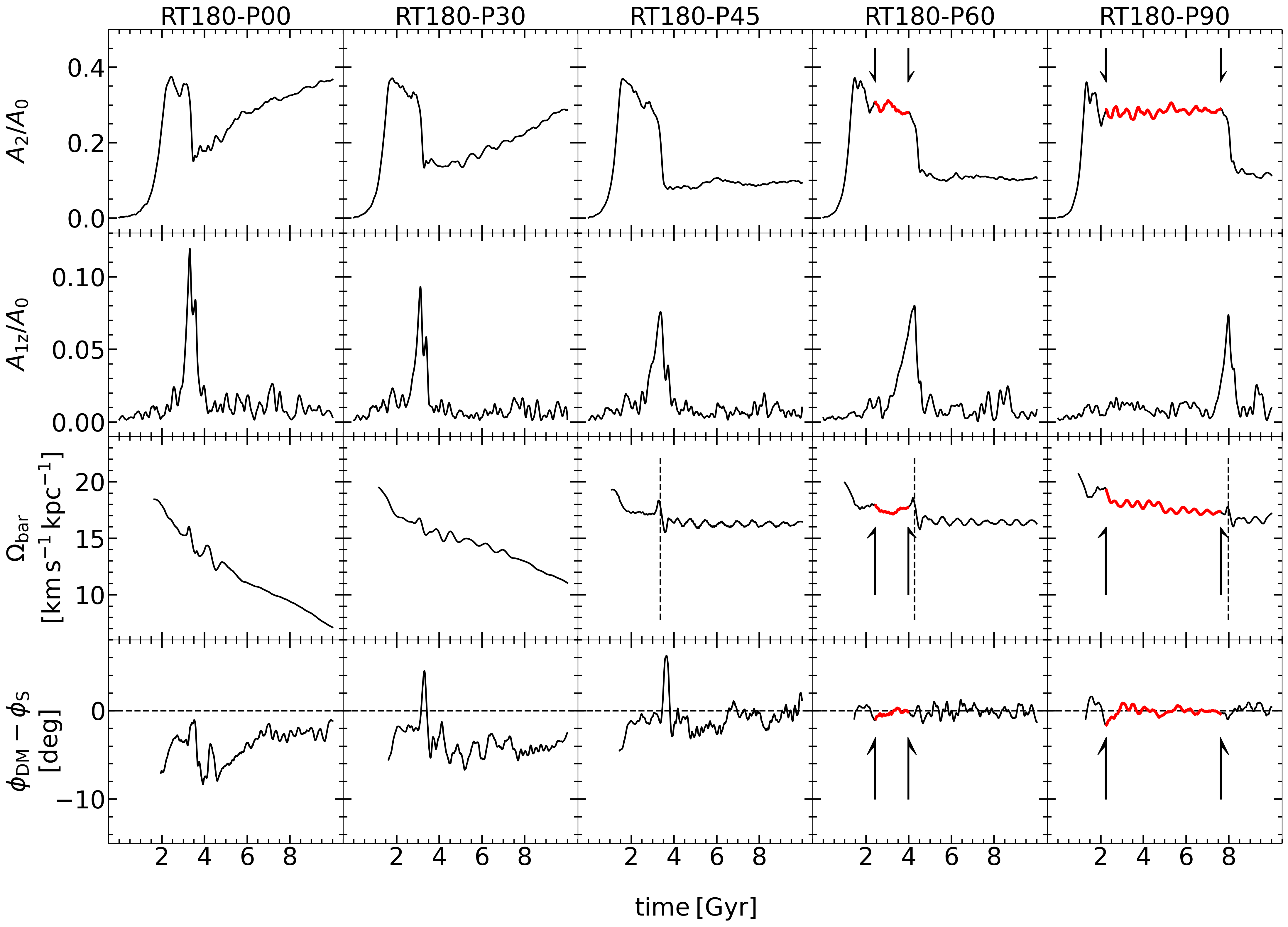}
    \caption{For RT180 halo model, normalized $\mathrm{A_2}$ (first row), pattern speed (second row), and normalized $\mathrm{A_{1z}}$ (third row) of the stellar bar , relative angle measured from halo two disc in the direction of the pattern speed (bottom row). The plateaus are colored red in between two arrows. The vertical dashed lines in the row of patter speed indicate the time of the maximum vertical buckling in some models. After the buckling, the stellar bars in these models dissolve and form the oval distortions.}
    \label{fig:rt180-all}
\end{figure*}

To verify the existence and properties of the plateau in the bar amplitude and its pattern speed, we focus on the Standard model, RT180, and supplement its $\uplambda$ sequence with additional models.  The $\uplambda$ sequence now consists of the following spin models: $\uplambda=0$, 0.03, 0.045, 0.06, and 0.09. Figure\,\ref{fig:rt180-all} displays the basic parameters of these models, namely, $A_2$, $\Omega_{\rm bar}$, $A_{1z}$, and the angle between the stellar and DM bars. 


The top row of Figure\,\ref{fig:rt180-all} exhibits the short duration plateau at $\uplambda=0.045$ with the length of about 0.6\,Gyr, and its increase to $\sim 2$\,Gyr for $\uplambda=0.06$, and to $\sim 5$\,Gyr at $\uplambda=0.09$. The bar pattern speed follows the same trend. During the plateau time, $\Omega_{\rm bar}$ stays nearly constant. The plateau is followed by the buckling and the bar dissolution. 

\begin{figure*}
\center 
	\includegraphics[width=0.8\textwidth]{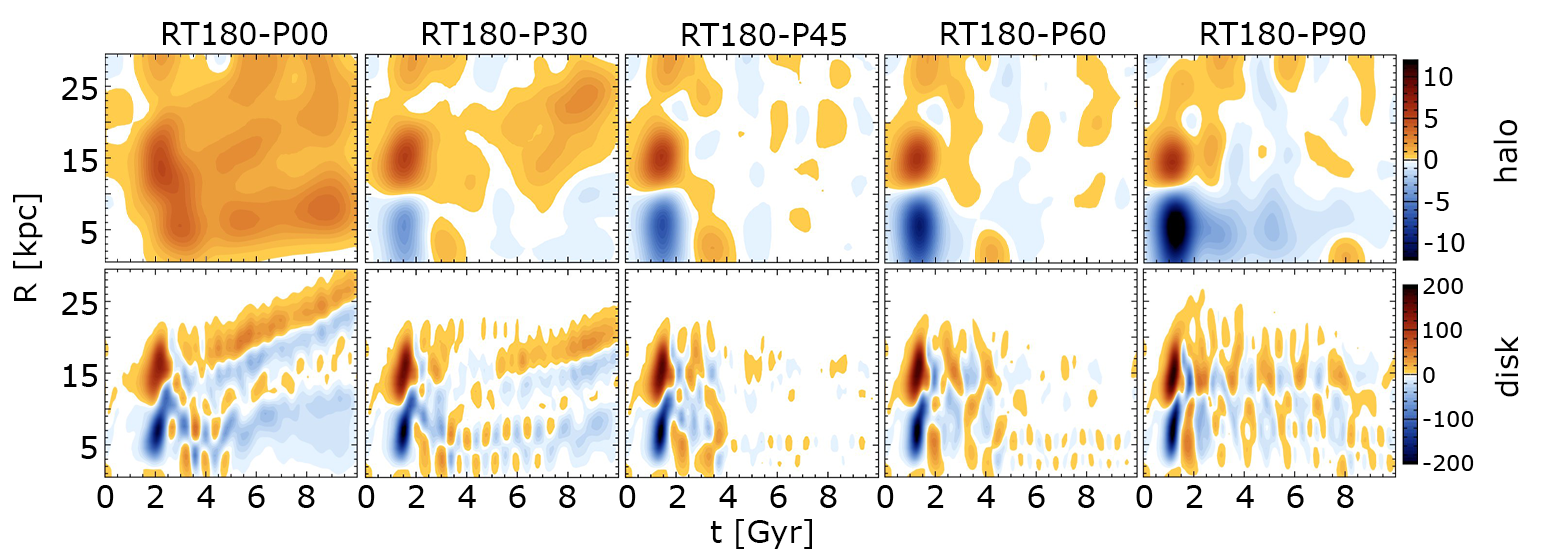}
    \caption{Same as Figure\,\ref{fig:jdot_rtseq_halo} but the halo (top row) and the disc (bottom row) in the RT-180 halo models.}
    \label{fig:jdot_rt180}
\end{figure*}

The bottom row of Figure\,\ref{fig:rt180-all} confirms that the DM and stellar bars are aligned during the plateau. Figure\,\ref{fig:jdot_rt180} exhibits the rate of the angular momentum transfer, $\dot J$, for the entire $\uplambda$-sequence of RT180.

\section{Discussion}
\label{sec:discuss}
 

We have performed numerical simulations of isolated disc galaxies embedded in DM halos. Two model sequences have been created --- varying the DM halo spin and varying the DM density. The DM density change has been achieved by keeping the halo mass fixed and varying its virial radius., i.e., by varying the halo compactness. The $\uplambda$-sequence confirms the results obtained in \citet{coll18} and \citet{coll19a}, while the RT-sequence reveals an interesting new trend in the bar evolution. We start with a short summary of our results and continue with their analysis.

\begin{itemize}
    
\item For DM halos with $\uplambda = 0$ and varying DM density, we detect no difference in their evolution --- bar strength, pattern speed, buckling time and its amplitude, etc. remain unchanged.

\item For the transition models with $\uplambda = 0.045$, decreasing DM density leads to a more complex behavior. The stellar bars reach their maximum strength earlier, but buckle later. The RT86P45 buckles at $t\sim 3.0$\,Gyr, RT150P45 at $t\sim 3.7$\,Gyr, and   RT210P45 at $t\sim 4.3$\,Gyr. This trend is not strong, the maximal delay is $\sim 1.3$\,Gyr, but we emphasize that $\uplambda = 0.045$ is a transition model. After buckling, the bar amplitude declines sharply, and regains strength progressively slower
with higher $\uplambda$. Starting with RT180 model and beyond, the bar does not regrow at all after buckling, leaving behind a weak oval distortion only.

\item The model sequence with $\uplambda = 0.06$, displays a more pronounced trend with a decreasing background DM density. The plateau appears for all the models except the P86 model, and the delay in buckling is pronounced.

\item For the $\uplambda = 0.09$ models with decreasing DM density, the behavior is even more extreme than in $\uplambda=0.06$. The delay in the buckling is more pronounced, and grows up to $\sim 5$\,Gyr.

\item Starting with RT150 model in the $\uplambda = 0.045$ sequence, and starting with RT118 model in the $\uplambda =0.06$ sequence, as well as for all the RT models in the $\uplambda=0.09$ sequence, the stellar bar strength remains constant after reaching maximum, at the level of $\sim 20\%$ below the maximum. In other words, the bar strength forms a plateau between the time of the maximum and the buckling. During this plateau, the stellar bars also exhibit only a very small decline in their pattern speed, despite being strong. A similarly reduced braking of stellar bar has been also reported by \citet{peter16} for a single model with $\uplambda = 0.03$ and by \citet{peter19} for $\uplambda = 0$. However, we cannot confirm this behavior for our models with $\uplambda \ltorder 0.04$.

\item The boxy/peanut shaped bulges appear at the beginning of the plateau, with no buckling instability observed. They grow monotonically along the plateau stage.  

\item For all our models the stellar bars lead the DM bars by an angle which decreases
{\it on the average} with $\uplambda$ and with decreasing DM density. This difference in the leading angle can be translated to gravitational torques exerted onto the stellar bar, and is discussed in the following.    
    
\end{itemize}

As a next step, we focus on the origin of the above plateau in some models of faster spinning halos with a lower DM density, and on the associated phenomena.

\begin{table}
	\centering
	\caption{{\it Left Column:} The $\uplambda$ sequence models. {\it Middle Column:} The angular momentum, $J_{\rm disc}$, lost by the disc during 10\,Gyr evolution, averaged over five RT models for each $\uplambda$. {\it Right Column:} Angles between the leading stellar bar and the trailing DM bar averaged over five RT models for each $\uplambda$ and over the evolution time. The time of buckling has been excluded from averaging.}
	\label{tab:table2}
	\begin{tabular}{lcc} 
		\hline
		Halo Spin & $\Delta J_{\rm disc}$ & $\phi_{\rm DM} - \phi_{\rm S}$ \\
                  & $\left[10^{10}\,\mathrm{M_\odot}\,\mathrm{kpc^2}\,\mathrm{Gyr^{-1}}\right]$ & degrees of arc\\ 
		\hline
		P00   & 2,751 & -6.0  \\
		  P45   & 1,060 & -2.2  \\
        P60   & 683   & -0.6 \\        
        P90   & 497   & -0.4  \\
		\hline
	\end{tabular}
\end{table}  

\subsection{Delayed buckling of stellar bars}
\label{sec:sumConclusions}

Our numerical simulations of stellar bar evolution in spinning DM halos have revealed an interesting effect --- reduction of the DM density in tandem with increasing spin resulted in the appearance of a plateau in the bar strength. The plateau has been achieved after the stellar bars reached their maximal strength. The bar remained strong during this time, only slightly below its maximum strength. The plateau time length has been found to depend on the DM halo spin $\uplambda$, and on the DM density --- its duration has been prolonged with the increasing halo spin and decreasing DM density (e.g., Figures\,\ref{fig:a2RT}--\ref{fig:omgBar-all} and \ref{fig:rt180-all}). The DM bars induced by the stellar bars are closely aligned with the stellar bar during the plateau. The stellar bars always buckled at the end of the plateau, and subsequently dissolved after buckling, leaving a weak oval behind.

The most intriguing property of stellar bars in the higher $\uplambda$ and higher RT sequences, i.e., lower density, models is the delayed buckling instability on timescales of $\sim 1-6$\,Gyr. This phenomenon appears to be associated with the formation of the plateau in the bar strength. During this time, the pattern speed of the bars remained nearly constant, and the stellar and DM bars remained aligned.

Table\,\ref{tab:table2} (middle column) provides the averages of the angular momentum loss, $\Delta J_{\rm disc}$, by the stellar discs during the 10\,Gyr of evolution, in the middle column. The stellar bars are responsible for the lion share of angular momentum loss by the discs. To observe the general trend, we have averaged the four representative $\uplambda = 0$, 0.045, 0.06, and 0.09 sequences over the RT models each. 

The P45 models average value of $\Delta J_{\rm disc}$ is about 2.6 times less than the P00 models. For P60 models this value is about 4 times smaller that for the P00 models. While P90 models average about 5.5 times less. Hence, increasing halo spin leads to a dramatic decrease of the angular momentum loss by the disc to the inner halo.

We have calculated the offset angles between stellar and DM bars and present their averages for each $\uplambda$ (Table\,\ref{tab:table2}, right column). These angles vary monotonically and are reduced from P00 to P90. The angle is reduced almost by a factor of 3 from P00 to P45, and by a factor of 15 from P00 to P90.  In fact, this angle is so small for P60 and P90, that one can conclude that the ars are basically aligned, thus minimizing the gravitational torques on the stellar bars. Altogether, this explains the reduced braking of stellar bars.

Note, that Figure\,\ref{fig:omgBar-all} shows that the braking is decreasing both by increasing $\uplambda$ and by decreasing the DM density, both horizontally (with the exception of P00 models) and vertically.  

The first question to ask is why the vertical buckling of the bar is delayed in some models presented here. Why does the bar strength stay constant for a long time, and why does the pattern speed of the bar remain nearly unchanged during the plateau stage? What determines the length of the plateau which is terminated by the buckling instability? We start by focusing on these questions, and discuss the corollaries of this evolution.

 \begin{figure}
\center 
	\includegraphics[width=0.4\textwidth]{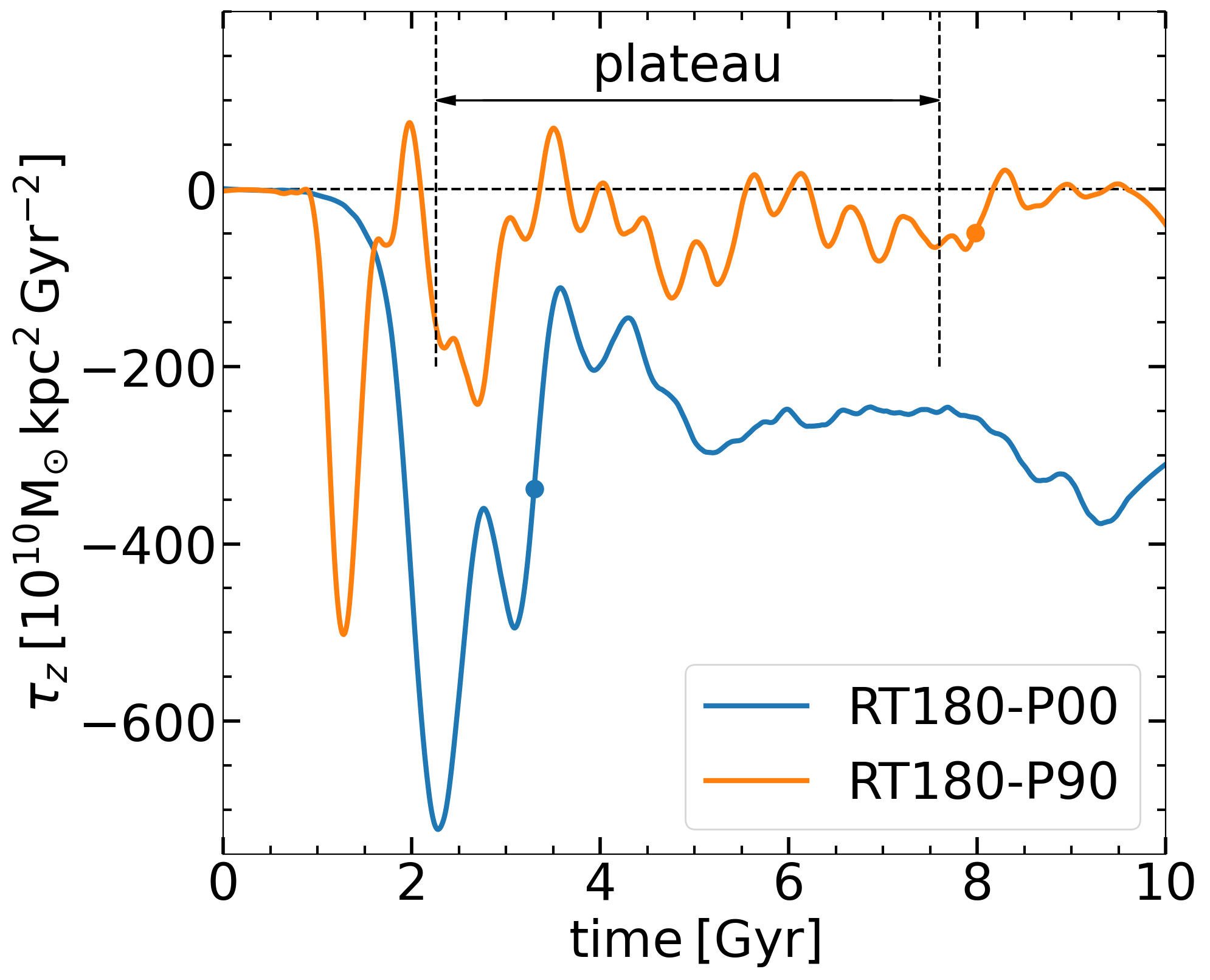}
    \caption{Evolution of the gravitational torques imposed by the DM on the stellar disc for the RT180 P00 and P90 models.
    The average torque during the plateau in the P90 model is about 4.8 times smaller than the torque imposed after buckling in the P00 model. The thick dots on the curves mark the maximal buckling stages of stellar bars in these models, and the horizontal arrow shows the extent of the plateau in the P90 model.}
    \label{fig:torque}
\end{figure} 

We have already determined in Figure\,\ref{fig:dphi_h2d_rt-all} that the angle between the stellar and DM bars depends inversely on $\uplambda$ and directly with the DM density, as shown in Table\,\ref{tab:table2} and Figure\,\ref{fig:rt180-all}. To verify this result, we have calculated the gravitational torques imposed by the DM on the stellar discs (Fig.\,\ref{fig:torque}) for our standard RT180 models P00 and P90. While the torque in P00 model has been calculated before \citep[e.g.,][]{bere07,villa10}, the torque on the disc immersed within the spinning DM halo is new. In the P00 model stellar bar, the torque it experiences is negative, i.e, it is braking. However, in P90 model, along the plateau, the torque is substantially smaller and oscillates around zero, with the average relative angle between the DM and stellar bars decreasing by a factor of 15 from P00 to P90 models.  Hence, the measures of torques and angles between DM and stellar bars confirm each other. Our next step is to analyze the physical processes occurring during the plateau stage in the P90 model.

To characterize the motions of stellar particles along the bar and along its rotation axis, we follow \citet{li23} and define the nonlinear frequencies, $\Omega_{x}$ and $\Omega_{z}$, where $\Omega_{x} = \Omega - \Omega_{\rm bar}$ is the radial frequency in the bar frame, and $\Omega_{z}$ is the oscillation frequency across the disc plane\footnote{We determine the period of the $i$-orbit as the time between two successive apocenters of this orbit, $T_{i}$, and their instantaneous frequencies as $\Omega_{x} = 2\pi /T_{i, x}$, for the motion along the bar major axis, and $\Omega_{z} = 2\pi /T_{i, z}$, for the vertical motion. See sections 5.1 and 5.2 in \citet{li23} for detailed discussion of how these frequencies have been calculated.}, i.e., along the $z$-axis. As noted in \citet{li23}, the resonances among these frequencies can lead to 2:1 and higher resonances, which will not be associated with the $\mathrm{x_2}$ family of orbits as defined by \citet{cont80}. Indeed, the $\mathrm{x_2}$ family has not been detected in our models. But resonances based on  $\Omega_{z}/\Omega_{x}$ being rational numbers have been detected. In this case, we did not call them the Lindblad resonances, but abbreviated them by the 2:1 ratios, and so for other resonances.

The advantage of using $\Omega_{x}$ and $\Omega_{z}$ frequencies is that the former one, $2\Omega_{x}$, acts as a driving force for vertical buckling of the bar, while the latter one acts as the natural vertical frequency of oscillators (i.e., of stellar particles). As discussed in \citet{li23}, when the ratio  $\Omega_{z}/2\Omega_{x} > 1$, the system of stellar particles can respond in tandem, while when $\Omega_{z}/\Omega_{x} < 1$, the response will be out of phase. This collective response will not be symmetric with respect to the stellar bar midplane, as internal stresses in the bar will be asymmetric. In the presence of self-gravity, the former case can lead to the buckling, while the latter case will not, and only result in heating up the stellar bar vertically. Of course, self-gravity is responsible for increasing the cohesiveness of the response. 

\begin{figure}
\center 
	\includegraphics[width=0.47\textwidth]{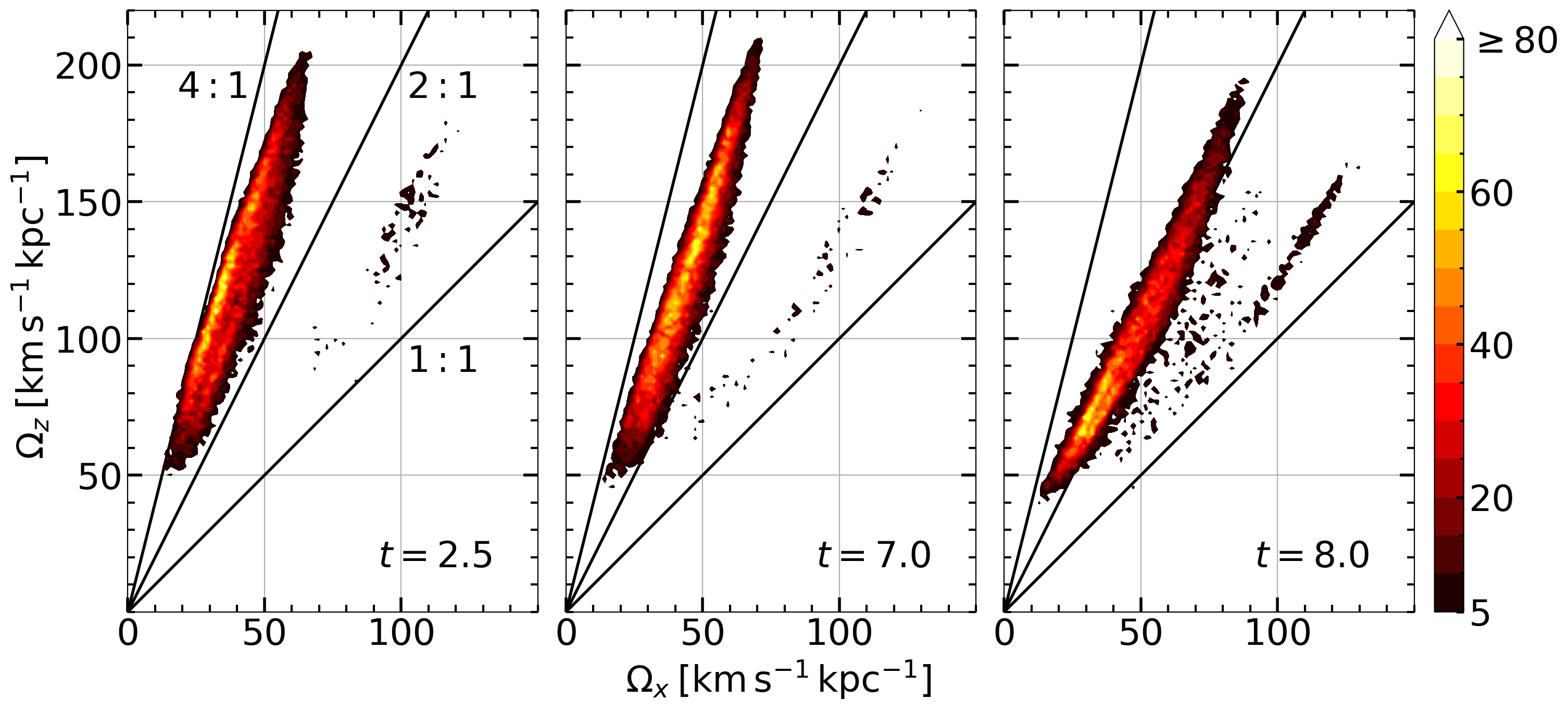}
    \caption{Evolution of the frequencies, $\Omega_{z}$ and $\Omega_{x}$, in the stellar bar during the plateau of the model RT180P90. note, the larger frequencies correspond to the innermost particles, and smaller frequencies refer to the outermost particles in the sample. The left snapshot represents the time close to the beginning of the plateau, $t\sim 2.5$\,Gyr, the middle snapshot is taken close to the end of the plateau, $t\sim 7.0$\,Gyr, and the right snapshot correspond to the maximal buckling of the bar. The sample of 30,000 particles has been selected randomly from the bar volume close to the maximum of the bar $A_2$ amplitude. Shown are the surface density contours (see the color palette for details). The diagonal lines represent (vertical) resonances in the  $xz$-plane: 1:1 resonance (the lowest line), 2:1 resonance (the middle line), and 4:1 resonance (the top line).  }
    \label{fig:OzOx}
\end{figure}

\begin{figure}
\center 
	\includegraphics[width=0.4\textwidth]{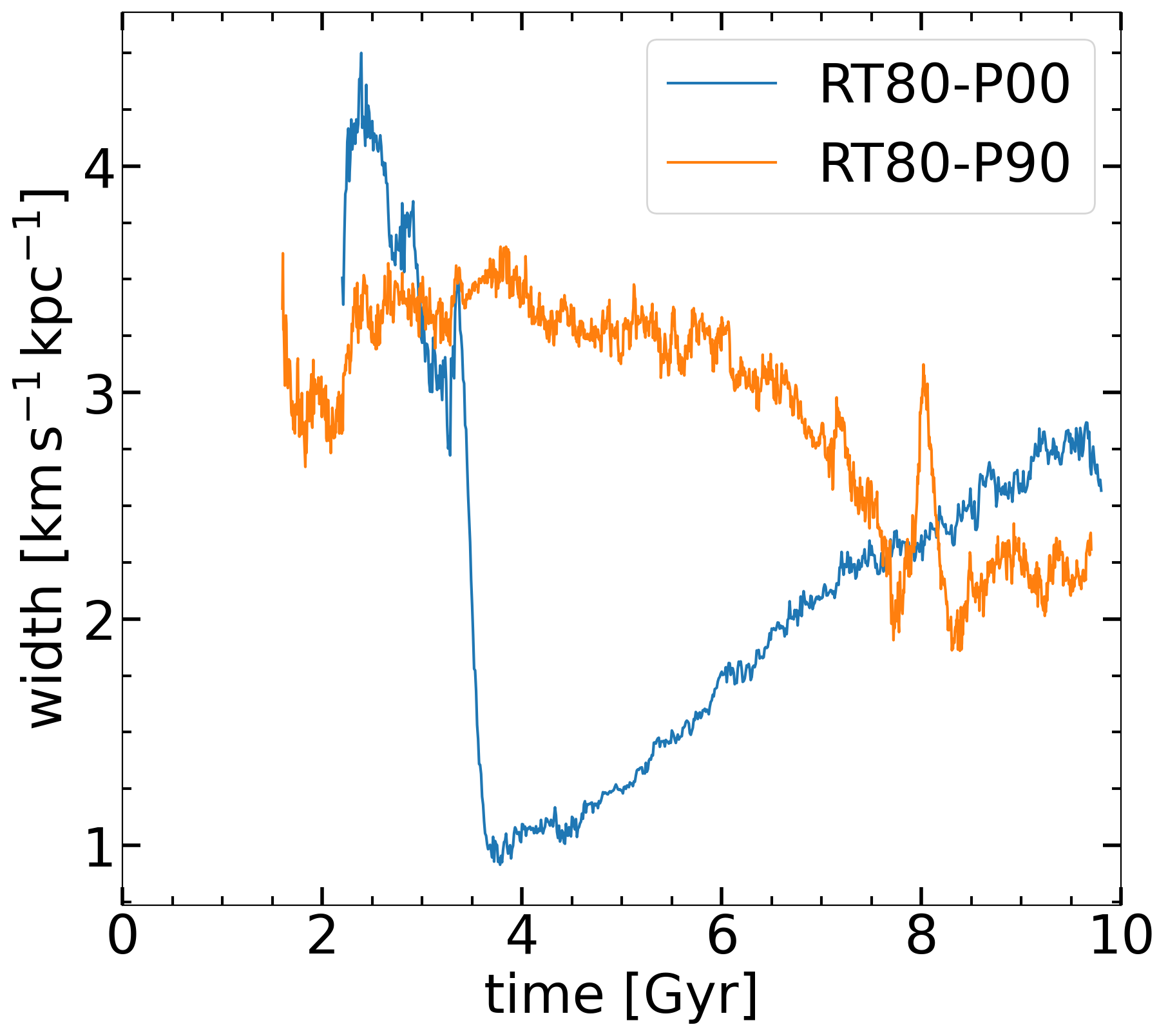}
    \caption{Evolution of the width of particle distribution in the $\Omega_{z}-\Omega_{x}$ plane shown in Figure\,\ref{fig:OzOx}. Only particles between the vertical 4:1 and 2:1 resonances are included.}
    \label{fig:d-width}
\end{figure}

\begin{figure}
\center 
    \includegraphics[width=0.4\textwidth]{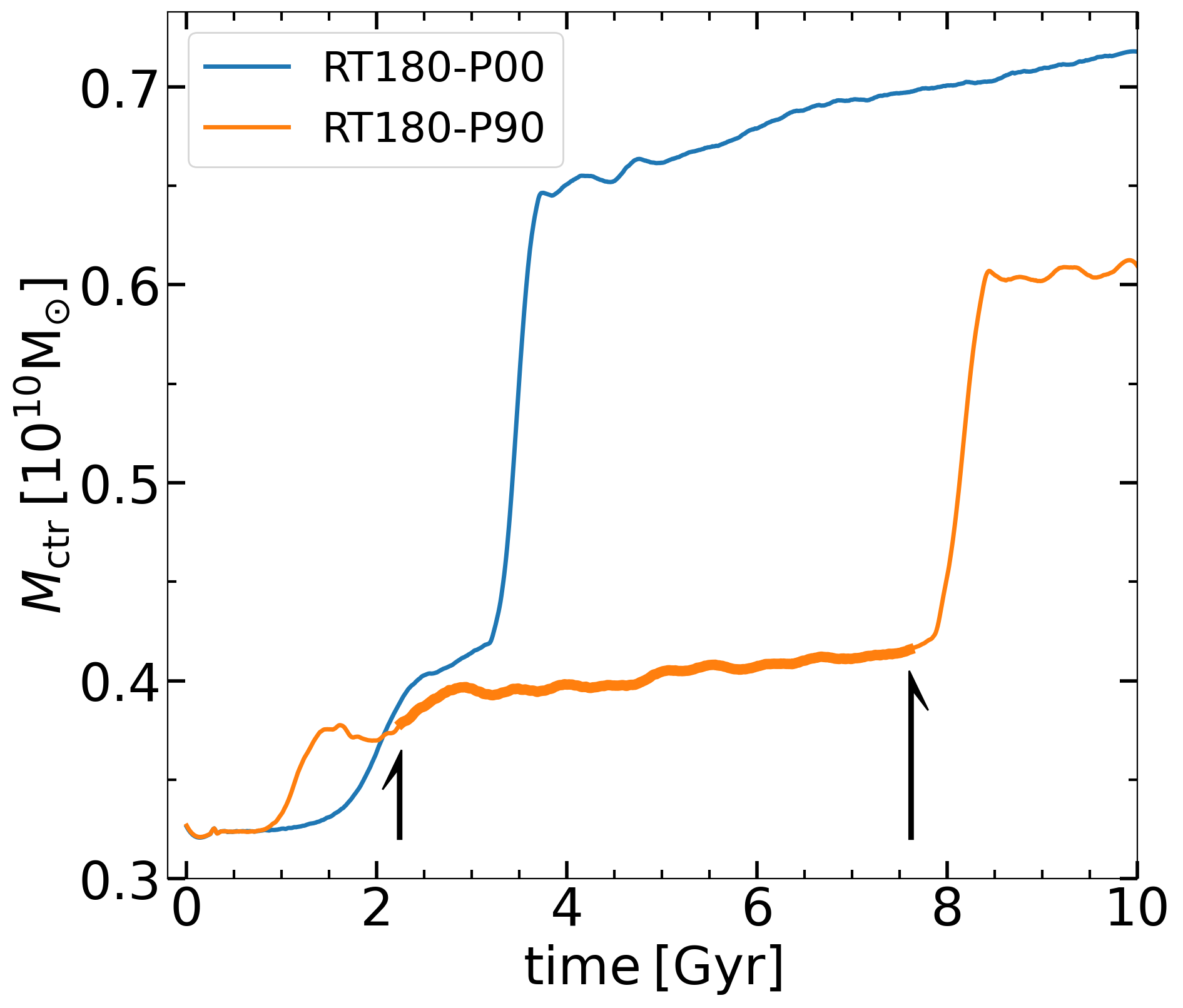}
    \caption{Evolution of the mass accumulation in the central region of the stellar bars in RT180 P00 and P90 models. The mass has been calculated within $|x| < 0.5$\,kpc, along the the bars, $|y| < 2$\,kpc perpendicular to the bars, and the vertical slice $|z| < 3$\,kpc. The P00 model buckles at $t\sim 3.3$\,Gyr, and P90 buckles at $t\sim 8$\,Gyr. The plateau in P90 model shown as the thick line in between two vertical arrows.}
    \label{fig:cmass}
\end{figure} 

To follow the evolution of stellar bars in the  $\Omega_{z}-\Omega_{x}$ frequency plane, we have adopted the method introduced by \citet{li23}, who applied it to nonrotating halos only. For the RT180P90 model, we randomly choose a sample of  30,000 stellar particles from the bar volume at the beginning of the plateau, $t\sim 2.5$\,Gyr, and calculate the $\Omega_{x}$ and $\Omega_{z}$ frequencies evolution along the plateau.

In the $\Omega_x - \Omega_z$ plane (Figure\,\ref{fig:OzOx}), we select a  \textit{main} population of stellar particles by choosing the particles in the two dimensional bins that contain more than 20 particles (the bin size is 1.5 by 1.5\,$\mathrm{km\,s^{-1}\,kpc^{-1}}$). We determine the width of the main population in this plane by applying the principal component analysis. Two principal components $\vec{p}_1$ and $\vec{p}_2$ have been obtained from the population. The first principal component, $\vec{p}_1$, is aligned with the long axis of the population in the above plane. The second principal component, $\vec{p}_2$, is orthogonal to the first one. Next, we transform the points $(\Omega_x, \Omega_z)$ of the main population into $(p_1, p_2)$ coordinates. The origin point of $p_1 - p_2$ space is taken to lie at the point of average of $\Omega_x$ and $\Omega_z$ of the main population, $(\left<\Omega_x\right>, \left<\Omega_z\right>)$. In the $p_1 - p_2$ space, we choose the particles with $|p_1| < 5$ and use the standard deviation of $p_2$ values from this sub-sample as the \emph{width} of the original main population in the $\Omega_{x} - \Omega_{z}$ plane.

As seen from Figure\,\ref{fig:OzOx} during the plateau, the sample distribution is elongated roughly along the vertical 4:1 resonance slope. The width of this distribution does evolve in the direction normal to the slope of the distribution. Figure\,\ref{fig:d-width} displays this evolution over the simulation time, starting with the bar formation phase. Initially, the distribution width increases, reaching the maximum around $t\sim 3.5$\,Gyr, then decreasing until $\sim 7.5$\,Gyr. Centered on $t\sim 8$\,Gyr --- the time of the vertical buckling, a sharp peak appears. What does evolution in Figures\,\ref{fig:OzOx} and \ref{fig:d-width} tell us?

The color in Figure\,\ref{fig:OzOx} describes the surface `density' in the frequency plane for stellar bar particles. The curve in Figure\,\ref{fig:d-width} follows the width of the distribution measured across the maximal surface density. With the bar reaching its maximal strength at $t\sim 1.4$\,Gyr and the beginning of the plateau at $\sim 2.3$\,Gyr, the width increases until $t\sim 3.5$\,Gyr. This time corresponds to the appearance of the peanut/boxy shaped bulge, which coincides with the vertical thickening of the bar, yet no buckling being observed. The subsequent decrease of the distribution width is related to the tightening of the distribution, measured perpendicularly to its slope. 

Along this direction normal to the slope, the surface density has a well-defined single peak. However, close to the buckling time, $t\sim 8$\,Gyr, a second peak appears making the distribution double-peaked, which results in a sharp increase of the distribution width and the appearance of a peak in Figure\,\ref{fig:d-width}. Importantly, this peak corresponds to the distribution crossing the vertical 2:1 resonance in the direction from the top down, as shown by the right snapshot of Figure\,\ref{fig:OzOx}. 

Hence, the crossing of the vertical 2:1 resonance by the distribution of stellar particles coincides with the appearance of the sharp peak in Figure\,\ref{fig:d-width}. In other words, the buckling happens when the distribution has a minimal width and, therefore, when the surface density in the frequency plane has reached the maximum. It is reasonable to assume that the cohesiveness of the particle distribution has reached a maximum at this time as well. Hence, this evolution hints at the plausible conditions suitable for the buckling.  
 
But why does the slope of the distribution remain nearly constant in time and only a short time preceding the buckling changes rapidly? \citet{li23}, in their Figure\,14, have shown that this slope changes due to the mass redistribution in the bar, i.e., sharply increasing the central density in the bar. This mass influx to the center generates circulation cells, i.e., increases vorticity there. We have tested this conclusion in Figure\,\ref{fig:cmass}, where we compared the evolution of the central mass accumulation in RT180P90 model with that of the RT180P00 one.

In both cases, we have observed the sharp increase in the central stellar mass at the time immediately preceding the buckling. Both models have evolved similarly in the narrow time interval surrounding the buckling instability. In both cases, this resulted in a rapid evolution of particles in the  $\Omega_{z}-\Omega_{x}$ frequency plane. The $\Omega_{x}$ frequency has increased while the $\Omega_{z}$ has decreased, which resulted in the particle distribution crossing the vertical 2:1 resonance slope.
 
\subsection{Particle trapping by planar and vertical 2:1 resonances}
\label{sec:trap}
 
\begin{figure}
\center 
	\includegraphics[width=0.4\textwidth]{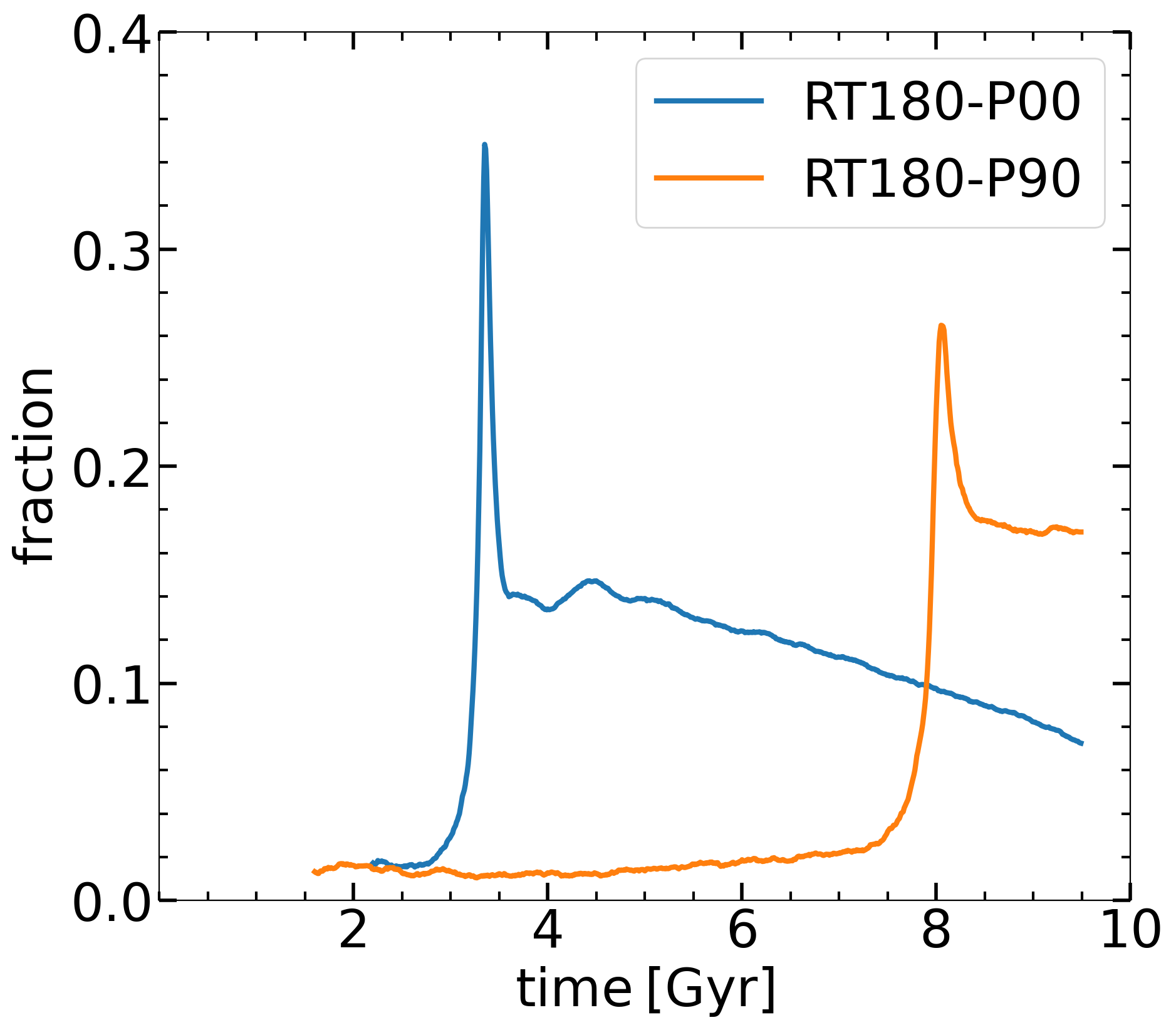}
    \caption{Evolution of the fraction of the bar particles that are swept by the vertical 2:1 resonance as a function of time, for $t\gtorder 2$\,Gyr for RT180 P00 and P90 models. The width of the resonance is taken assuming $\Delta_{\rm z} = \Delta_{\rm x} = \pm 5\,{\rm km\,s^{-1}\,kpc^{-1}}$. }
    \label{fig:swept}
\end{figure}

As a next step, we have determined the trapping of stellar particles in the bar by the vertical 2:1 resonance, as they are swept by this resonance in the RT180P90 model, and compared it with the RT180P00 model from  \citet{li23}. Figure \,\ref{fig:swept} exhibits the fraction of trapped particles assuming the resonance width of $\pm 5\,{\rm km\,s^{-1}\,kpc^{-1}}$.  The respective curves peak strongly at the time of the maximum of buckling amplitude, with a fraction $\sim 35\%$  and $\sim 27\%$ associated with each peak. Li et al. have argued that this fraction is sufficient  to assure the cohesiveness of the buckling (i.e., collective) response, and pointed to the important role of resonances in the buckling instability. 

\begin{figure}
\center 
	\includegraphics[width=0.4\textwidth]{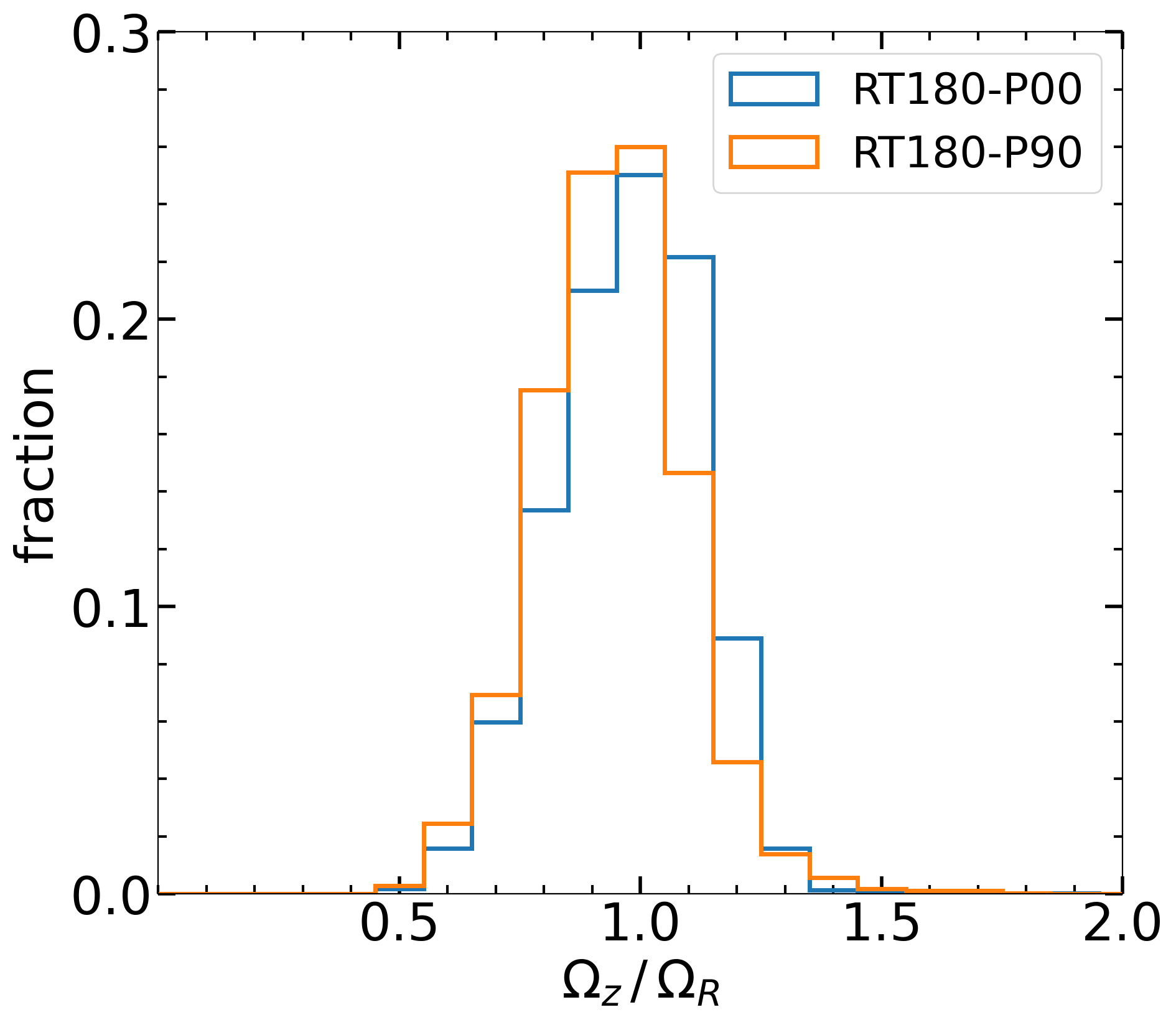}
    \caption{Fraction of sampled stellar bar particles normalized by the total number of particles in the sample as a function of $\Omega_{\rm z}/\Omega_{\rm x}$ ratio for two models, RT180P00 and RT180P90. This Figure shows the overlapping planar and vertical 2:1 resonances at the time of maximum of the vertical buckling amplitude, i.e., at $t\sim 3.33$\,Gyr and $\sim 8.05$\,Gyr, respectively. More than a quarter of the particles trapped in the planar resonance are trapped by the vertical resonance. The width of the resonance is taken assuming $\Delta_{\rm z} = \Delta_{\rm x} = \pm 5\,{\rm km\,s^{-1}\,kpc^{-1}}$. }
    \label{fig:overlap}
\end{figure} 

Simultaneous trapping by planar and vertical resonances provides an important signature of coupling between the $xy$- and $xz$-planar motions of stellar particles. To analyze the overlap between the planar and vertical 2:1 resonances, we have defined the nonlinear planar, i.e., $xy$, 2:1 resonance as $\Omega_{\rm R}$. This frequency is a nonlinear extension of the epicyclic frequency $\kappa$. The planar 2:1 resonance occurs when $\Omega_{\rm R}/\Omega_{x} = 2$.
 
Figure\,\ref{fig:overlap} displays this overlap between the planar and vertical 2:1 resonances for RT180 P00 and P90 models. The fraction of the particles with $\Omega_{z}/\Omega_{\rm R} = 1$ peaks at the maximal buckling moments for each of the above models. More than a quarter of orbits of stellar particles are trapped by these resonances at this moment.
 
The buckling instability is associated with an abrupt growth of the central mass accumulation in stellar bars and particular velocity field with an elevated vorticity \citep{li23}. We have tested these conclusions for the models with spinning halos which display the same abrupt mass increase for our standard models RT180P90 in comparison with RT180P00 (Fig.\,\ref{fig:cmass}). During the plateau stage in the former model, the central mass increases very slowly followed by a sudden increase during buckling. Note that the runaway increase in the central mass occurs at the same value in both models. However, it is not clear whether this conclusion is also valid for different disc and halo mass distributions.  
 
\subsection{Corollaries for the plateau in $\Omega_{\rm bar}$, $A_2$ and ${\rm R_{bar}}$}
\label{sec:observ}
 
We have shown that evolution of stellar bars depends both on the parent DM halo spin and its DM density. This evolution differs dramatically from that inside nonspinning halos --- an effect that has substantial observational corollaries. The apparent discrepancy between the observed pattern speeds of stellar bars, when using the Tremaine-Weinberg method \citep{tremaine84}, and those detected in the pure $N$-body modeling of disc-halo systems \citep[e.g.,][]{atha03,marti06,villa09} is difficult to reconcile without invoking the DM halo spin and dependency on the DM density. Many observed bars exhibit substantially higher pattern speeds compared to those in isolated collisionless models in numerical simulations.

Two possible explanations emerge of why the observed stellar bar pattern speeds appear so diverse \citep[e.g.,][]{deba02,laine98,guo19,garma20}. The first option involves the presence of gas which exerts a positive torque on the stellar bar, and which can lead to sharply reduced {\it net} torques on the stellar bar --- basically a plateau in $\Omega_{\rm bar}$, $A_2$ and in the bar size, $\mathrm{R_{bar}}$, as shown in numerical simulations of isolated \citep{bere07,villa10} and cosmological \citep{frag21} disks, especially in the gas-rich galaxies at high redshift \citep{bi22}. The origin of this plateau differs fundamentally from the plateau discussed in this work. The second option has been analyzed here in a sequence of models, and shows that stellar bars in spinning halos with lower DM density experience a dramatically weakening brake, because of the reduced gravitational torques from the DM, and maintain high $\Omega_{\rm bar}$ and fixed bar size even in the absence of a gaseous component while the stellar bar itself remains strong.

The presence of stellar bulges in disk galaxies is known to affect the evolution of stellar bars, including their vertical buckling process, and the evolution of the bar strength \citep[e.g.,][]{fujii18, sell20}.

Interestingly, a single model with $\uplambda = 0.03$ and a cuspy halo has shown a reduced braking of the bar by a factor of $\sim 3$ when compared to the model with an axisymmetrized halo for the duration of 2.4\,Gyr \citep{peter16}, although these models are different. No plateau in $A_2$ and the bar size have been mentioned, and the buckling happened at $\sim 2$\,Gyr, later compared to the similar nonrotating halo which buckles at 1.2\,Gyr. Moreover, \citet{peter19} analyzed two models with cuspy and cored DM halos with $\uplambda = 0$, and found a plateau in both models in $\Omega_{\rm bar}$ and bar size. In the core model, the plateau happens for about 1.2\,Gyr before the subsequent growth of the stellar bar continues. In the cuspy model, the plateau happens after the bar growth stage, for about 2.6\,Gyr.

Our models with $\uplambda \ltorder 0.04$ do not show the behavior encountered by \citet{peter19}. Yet, we have tested our models for $\uplambda = 0$, 0.015, 0.03, 0.045, 0.06, 0.075 and 0.09 \citep{coll18}, as well as for a range in the DM densities (this work). In particular, our plateau stages in $A_2$ and $\Omega_{\rm bar}$ and ${\rm R_{bar}}$ always end up with the buckling. The discrepancy between these simulations should be resolved in future work. 

Our choice of $\uplambda = 0 - 0.09$ encompasses the lion share of DM halos in the lognormal $\uplambda$ distribution expected in the universe. \citet{bull01} obtained $\overline\uplambda = 0.035$, while \citet{jiang19} provides a mean value of $\overline\uplambda = 0.037$. However, additional factors can play a role in this value, such as the environment the halos are found, e.g., in cosmological filaments or in voids \citep[e.g.,][]{Veena18}.

Based on the lognormal distribution of $\uplambda$ found by \citet{bull01}, about 30.8\% DM halos have $\uplambda \gtorder 0.045$. These halos should substantially modify the evolution of stellar bars in disk galaxies. Of course, if the inner halos have an excess of the spin over the universal angular momentum distribution in the halo \citep[e.g.,][]{bull01}, this will reduce the limits on the spin which affects the bar evolution and will increase the fraction of halos which are involved in this process. 

Moreover, the first attempt to estimate the spin of the DM halo for the Milky Way galaxy has resulted in the range of $\uplambda \sim 0.061-0.088$, meaning that 21\% of DM halos have $\uplambda \gtorder 0.06$ \citep{obreja22}. This high value may hint that additional effects, like baryons, can increase the spin of the inner halo, which is most relevant for the effect analyzed in this paper.

The effect discussed in this work allows to suggest that the diversity of observed pattern speeds in stellar bars underlines the distribution of DM halo spins and their DM densities, and the ability of strong bars to avoid a rapid braking if residing in the higher spin and lower DM halos.

\section{Conclusions}
\label{sec:conclusion}

We presented a suite of numerical simulations of disc-halo models subject to stellar bar instability followed by the vertical buckling instability. The DM halos have been designed to form two sequences along which the halo spin, $\uplambda\sim 0-0.09$, and concentration, or halo size, have been modified. All the halos have been of the same total masses and the NFW mass distribution with identical embedded stellar discs.

Our main result is that evolution of stellar bars in spinning DM halos is sensitive to the DM density, and this effect is strongly amplified with increasing DM halo spin, in contrast with the nonrotating halos. We have demonstrated that for $\uplambda\gtorder 0.045$ and lower DM density the buckling instability of the stellar bars begins to be gradually postponed after the bar has reached its maximal strength, by $\Delta t$ up to $\sim 5$\,Gyr. For a fixed $\uplambda$, the effect increases with decreasing DM density. Models with $\uplambda\sim 0.045$ appear to be a transition sequence.

We find that the boxy/peanut shaped bulges in the models with delayed buckling form after the bar has reached its maximal strength. and grow secularly until the buckling develops. This further confirms and demonstrates that buckling is not a requirement for a boxy/peanut bulge formation --- it can form either dynamically during buckling, or secularly and unrelated to buckling.

Moreover, the stellar bars embedded within higher DM spin halos show progressively smaller offset angle with the trailing DM bars, thus reducing the gravitational torques between them. As a result, the pattern speeds of stellar bars remain nearly constant for time periods of up to $\sim 5$\,Gyr. This plateau in the stellar bar pattern speed occurs in tandem with the bar exhibiting nearly maximal strength and unchanged bar size.

The above phenomena, the plateau in the bar strength and size, and near absence of the stellar bar braking in  pure $N$-body systems have been observed before in \citet{peter19}, but for a model of a {\it nonrotating} halo. Although both our and Petersen et al. works are based on the orbit analysis of numerical simulations, we could not confirm this result which must be reconciled in future numerical simulations.

The observed discrepancy between observed faster pattern speeds of stellar bars and their modeled slower counterparts can be explained either by positive torques applied by the gas on the stellar bar, or by stellar bars residing in the faster spinning DM halos which lead to the orbit trapping and reduced gravitational torques. The presence of a gaseous component of $\gtorder 5\%$ has been modeled and explained by the positive torque applied on the bar by the gas in nonspinning DM halos \cite{bere07,villa10}. However, in our present simulations, the gas component is absent and the reported action comes solely from the spinning parent DM halo. 

During the plateau, stellar bars do not grow in size and their corotation radii do not increase as well. However, the reduced density leads to pronounced changes in the models with the spinning halos. The characteristic frequencies of stellar oscillations along the major axis of the bar and its vertical axis experience a very slow evolution during the plateau stage. We have performed the orbit analysis for the disc-halo systems along the $\uplambda$ and DM density sequences and found that the ratio of these frequencies which define the vertical 2:1 resonance is reached much later, at the end of the plateau stage. This ratio is governed by the mass distribution within the stellar bar. The mass distribution is changing very slowly during the plateau stage, but increases abruptly during the buckling stage. As the central mass distribution in the stellar bar governs the appearance of the vertical 2:1 resonance and hence of the onset of the vertical buckling instability in the stellar bar, this instability is delayed. During buckling, a substantial fraction of stellar particles in the bar is trapped by the overlapping 2:1 vertical and planar resonances, thus enhancing the rate of energy transfer from the planar to vertical motions in the bar, in agreement with \citet{li23}.

Evolution of stellar bars in spinning halos with various DM densities appears to display a rich family of possible solutions, hinting at additional phenomena to be discovered and analyzed before the subject of the $N$-body dynamics is closed.  
 
\section*{Acknowledgements}

We thank Phil Hopkins for providing us with the latest version of GIZMO and Angela Collier for sharing some of the analysis software. I.S. is grateful for a generous support from the International Joint Research Promotion Program at Osaka University. This work has also been supported in part by the Hubble Theory grant HST-AR-18584, by JSPS KAKENHI grant 16H02163 (to I.S.) and by the NSF under Grant PHY-1748958 to Kavli Institute for Theoretical Physics (KITP).  The STScI is operated by the AURA, Inc., under NASA contract NAS5-26555. Simulations have been performed using the University of Kentucky Lipscomb Computing Cluster. We are grateful for help by Vikram Gazula at the Center for Computational Studies at the University of Kentucky for help with the technical issues with the LCC runs.

\section*{Data Availability}

The data presented in this work can be obtained upon reasonable request.



\bibliographystyle{mnras}
\bibliography{paper} 

 









\bsp	
\label{lastpage}
\end{document}